\documentclass[aps,prb,10pt,superscriptaddress,twocolumn]{revtex4-2}

\usepackage[pdftex]{graphicx}
\usepackage{dcolumn}
\usepackage{bm}
\usepackage{color}
\usepackage{amsmath}
\usepackage{nicefrac}
\usepackage{amssymb}
\usepackage[bb=boondox]{mathalfa}

\usepackage[dvipsnames]{xcolor}
\usepackage{dsfont} 
\usepackage{braket}
\usepackage{tensor}
\usepackage{mathtools}
\usepackage[caption=false]{subfig}
\usepackage{utfsym}  

\newcommand{\ff}[1]{{\boldsymbol #1}}
\newcommand{\ca}[1]{{\cal #1}}
\newcommand{\bi}{\begin{itemize}}
\newcommand{\ei}{\end{itemize}}
\newcommand{\be}{\begin{equation}}
\newcommand{\ee}{\end{equation}}
\newcommand{\ba}{\begin{eqnarray}}
\newcommand{\ea}{\end{eqnarray}}

\usepackage[pdftex,colorlinks=true,linkcolor=blue,citecolor=magenta,filecolor=blue]{hyperref}

\newlength{\minuswidth}
\newcommand{\makeupminus}[1]{\hspace*{0.5\minuswidth}#1\hspace*{0.5\minuswidth}}

\begin{document} 
	
\title{Exchangeless braiding of Majorana zero modes in weakly coupled Kitaev chains}
	
\author{Robin Quade}
	
\affiliation{I. Institute of Theoretical Physics, Department of Physics, University of Hamburg, Notkestra\ss{}e 9-11, 22607 Hamburg, Germany}
	
\author{Michael Potthoff}
	
\affiliation{I. Institute of Theoretical Physics, Department of Physics, University of Hamburg, Notkestra\ss{}e 9-11, 22607 Hamburg, Germany}
	
\affiliation{The Hamburg Centre for Ultrafast Imaging, Luruper Chaussee 149, 22761 Hamburg, Germany}
	
\begin{abstract}
Exchangeless braiding of Majorana modes is studied in minimal networks of weakly hybridized Kitaev chains of finite length using a rigorous many-body framework.
In particular, for two coupled chains it is shown that exchangeless braiding is achieved by $2\pi$ rotations of the phase $\phi$ of the superconducting order parameter of one of the chains. 
This braiding protocol is verified by the numerical calculation of the non-Abelian Wilczek-Zee phase of the low-energy many-body subspaces $\ca H_{0}(\phi)$ based on the Bertsch-Robledo ground-state overlap formula.
In the parameter space spanned by the total chain length, the strength of the weak hybridization connecting the chains, and the on-site potential, we identify two regions with different braiding outcomes, i.e., a projective $\sigma_{x}$-gate and a projective $\sigma_{z}$-gate phase.
The transition between these phases is a continuous crossover, the location of which is reliably given by a simple four-Majorana mode model.
This demonstrates the resilience of the anyonic properties of the Majorana modes against finite-size effects and weak links between Kitaev chains.
\end{abstract} 
		
\maketitle 
	
\section{Introduction}
\label{sec:intro}

Topological quantum computation (TQC) is one of the most promising approaches toward fault-tolerant quantum computation (QC) \cite{Kitaev2003,Nayak2008,Moore1991,freedman2002,Nayak1996,Freedman1998,DasSarma2005,Bonderson2008,Pachos2017,DasSarma2015,Pachos2012,DasSarma2006,Collins2006}. 
It is based on the exotic statistics of non-Abelian anyons, which provides topological protection against local perturbations and control errors  \cite{Kitaev2003,Nayak2008,Witten1989,Leinaas1977,Rao2017}. 
More precisely, in certain quantum systems, the presence of anyonic quasi-particles gives rise to a degenerate low-energy subspace $\mathcal{H}_{0}$ suitable for encoding quantum information \cite{Pachos2017}. 
The adiabatic exchange of anyonic quasi-particles induces unitary transformations on $\mathcal{H}_{0}$ that depend only on the topology of the braids formed by the anyonic world lines during the exchange. 
If $\mathcal{H}_{0}$ is used to encode a qubit $\ket{\psi}$, protocols of adiabatic anyon exchanges can be applied to implement robust quantum gates on $\ket{\psi}$. 
As a result, TQC with anyons implements error correction on a hardware level, making it a compelling architecture for practical QC applications.
	
In the recent decades, TQC has attracted increasing interest, driven in part by the discovery of anyonic defect modes in certain topological superconductors  \cite{Ivanov2001,Kitaev2001,Read2000,Fu2008,Alicea2012,Alicea2011,Chiu2016}. 
These defect modes are exotic in several respects. 
First, their existence is ensured by the bulk-defect correspondence of the underlying topological superconductor, which protects them against symmetry-preserving perturbations and locks them at zero energy \cite{Chiu2016}. 
The latter leads to a degeneracy of the many-body ground-state energy, introducing the degenerate subspace $\mathcal{H}_{0}$ of ground states needed for QC. 
Second, the associated zero-energy quasi-particle excitations of the superconducting state can be understood as self-adjoint quasi-particle modes, i.e., Majorana-type fermions. 
For this reason, they are usually referred to as Majorana zero modes (MZMs). 
Finally, and most importantly for TQC, they are equivalent to a particular class of non-Abelian anyons, known as Ising anyons \cite{Pachos2017,Rao2017,Nayak2008}.
Even though Ising anyons are not universal for quantum computing in that not every unitary transformation on $\mathcal{H}_{0}$ can be approximated to arbitrary precision by braiding transformations,
their experimental accessibility and low braiding complexity make the study of physical Ising anyons a worthwhile endeavor.

Proposals for concrete physical realizations include MZMs bound to vortex cores in two-dimensional $p_{x} + i p_{y}$ superconductors \cite{Ivanov2001,Read2000,Tewari2007,Sanno2021} and boundaries of one-dimensional topological $p$-wave superconductors \cite{Kitaev2001,Oreg2010,Yuval2010,Alicea2011,Pientka2013,Halperin2012,Kraus2013,Amorim2015,Sekania2017,Aasen2016,Karzig2017,Harper2019,Tutschku2020,Tanaka2022}. To explore these physical implementations, several experimental and theoretical challenges must be overcome. Experimentally, there is an ongoing effort to develop engineering techniques for topological superconductors and to achieve the microscopic control required for the preparation and manipulation of MZMs, see Refs.\ \cite{Flensberg2021,Li2014,NadjPerge2014,Ruby2015,Jeon2017,Pawlak2016,Krogstrup2015,Chang2015,Sestoft2018} for examples.
On the theoretical side, it is necessary to incorporate more real-world constraints into simulations in order to arrive at reasonable expectations for experimental implementations. 
Progress has been made with several methodological approaches. 
Exact diagonalization is effective but limited to small systems \cite{Sekania2017,Li2016,Wieckowski2020}, while studies focusing on single-particle states \cite{Amorim2015,Tutschku2020} or low-energy theories \cite{Karzig2013,Chen2022,Karzig2015,Knapp2016} can achieve larger system sizes at the cost of neglecting bulk effects. 
Other efforts include time-evolving quasi-particles \cite{AEY+15,Cheng2011,Scheurer2013,Sanno2021,ZDS+22}, analyses based on the Onishi determinant formula \cite{Conlon2019}, and approaches using the covariance matrix \cite{bravyi2012,Bauer2018,Truong2023}. 
	
The present work reports many-body simulations for weakly coupled, finite-size Kitaev chains.
We suggest an exchangeless braiding protocol, in which the unitary braiding transformation is driven not by an adiabatic exchange of MZMs, but by $2\pi$ rotations of the superconducting phase $\phi$ that keep them stationary in real space \cite{Bonderson2008,Bonderson2009,vanHeck2012,Burrello2013,Hyart2013}. 
This was originally proposed by Kitaev \cite{Kitaev2001} and later picked up by others, including Fu, Teo and Kane \cite{Kane2008,TeoKane2010}, Chiu et al.\ \cite{Chiu2016}, and Sanno et al.\ \cite{Sanno2021}. 
The Kitaev chain model \cite{Kitaev2001} recently experienced a considerable renewal of general interest because it was found to provide a suitable low-energy description of some semi-conductor/superconductor or spin-chain/superconductor hybrid systems \cite{Tewari2007,Pientka2013}. 

Local manipulations of the superconducting phase are generally difficult to realize. 
Here, their use is motivated by the observation that in Yu-Shiba-Rusinov (YSR) platforms for topological superconductivity \cite{Pientka2013,NadjPerge2013,Braunecker2010,Choy2011,Kjaergaard2012,Martin2012}, an {\em en bloc} rotation of the magnetization direction of a ferromagnetically aligned classical spin array exchange-coupled to a conventional $s$-wave superconductor translates into a phase rotation of the proximity induced $p$-wave superconducting gap in the YSR bands, as discussed in Ref.\ \cite{Pientka2013}. 

To detect the presence or absence of Ising MZM braiding, we analyze the non-Abelian Wilczek-Zee (WZ) phase \cite{Wilczek1984} of the low-energy subspace $\mathcal{H}_{0}(\phi)$ under the cyclic evolutions $\phi \mapsto \phi + 2\pi$ of the superconducting phase. 
Thus, we explicitly evaluate a $\text{U}(n)$ holonomy associated to the $n$-dimensional space $\mathcal{H}_{0}(\phi)$ focusing our considerations to the geometric while disregarding the dynamical contributions.

This is justified in the physically relevant limit, where the superconducting gap is much larger than the inverse $\tau^{-1}$ of the typical time scale of the real-time dynamics, and where $\tau^{-1}$ is much larger than a slight energy splitting of the MZMs. 
In this case the dynamical phase factor is just an overall U(1) phase and can be disregarded, such that the dynamics is fully described by the geometrical Wilczek-Zee phase matrix.

Methodologically, we follow Ref.\ \cite{Mascot2023} and use the Robledo-Bertsch-Pfaffian overlap formula \cite{Robledo1994,Robledo2009,Robledo2011,Bertsch2012,Neergard2023} to circumvent the Onishi sign problem in the computation of many-body overlaps between zero-energy (low-energy) Fock states in the numerical computation of the WZ phases.
The geometrical Wilczek-Zee phase matrix relevant for MZM braiding must be computed from the $\phi$-dependent many-body ground state $| 0 \rangle_{\rm b}$, the Bogoliubov vacuum, and further states with a single, $b^{\dagger}_{m} | 0\rangle_{\rm b}$, or more low-energy Bogoliubov quasi-particle excitations. 
While the $\phi$-dependence of $b_{m}^{\dagger}$ is accessible, e.g., within time-dependent Bogoliubov-de Gennes theory, the present Fock-space approach additionally accounts for the non-trivial evolution of $| 0 \rangle_{\rm b}$ under variation of $\phi$. 
This characteristic property of the ground state of a superconductor and the resulting difficulties that are fully solved with the Fock-space approach have been summarized by Mascot et al.\ \cite{Mascot2023}.

We start our study with a numerical analysis of the WZ phase for a single Kitaev chain.
It is demonstrated that the MZMs retain their anyonic properties even for rather short chains, although their energy degeneracy is lifted significantly.
This stabilitiy enables the implementation of an exchangeless double-braiding protocol driven by cyclic rotations of the superconducting phase. 
Our main focus is on two weakly linked chains. 
Here, we find that the relative strength of the inter- and intra-subchain coupling between different MZMs can be used to control, if an exchangeless double braiding of the MZMs of either subchain results in a unitary $Z$- or $X$-gate transformation.

The paper is organized as follows. Sections \ref{sec:mod} and \ref{sec:meth} introduce the model and outline the technical framework underlying our numerical approach.
In section \ref{sec:one_KC} we numerically evaluate the Wilczek-Zee phase to demonstrate exchangeless double braiding and quantify finite-size effects in a single Kitaev chain. Section \ref{sec:two_KC} extends this analysis to a simple two-Kitaev chain network, where the exchangeless double braiding process is shown to induce $Z$- or $X$-gate transformations, depending on the parameters of the model.
Section \ref{sec:conclusion} concludes with a summarizing discussion.

\section{Kitaev Chain}
\label{sec:mod}

Our numerical computations have been performed for the Kitaev chain \cite{Kitaev2001} and variants of it. 
Here, we briefly introduce the Kitaev-chain model and fix some notations that will be used later.

Using standard notations with annihilation and creation operators $c_{j}$ and $c_{j}^{\dagger}$, the Hamiltonian is given by 
\ba
H_{\text{Kit}} &=& \sum_{j = 1}^{L}  \Big[ -t \, (c_{j+1}^{\dagger} c_{j} + c_{j}^{\dagger}c_{j+1}) - \mu \Big(c_{j}^{\dagger}c_{j} - \frac{1}{2} \Big) \nonumber \\
&+& \Delta \, c_{j}c_{j+1} + \Delta^{\ast} \, c_{j+1}^{\dagger}c_{j}^{\dagger}\Big]  
\label{H_Kit}
\ea
and describes a system of spinless itinerant fermions hopping over the sites $j=1,...,L$ of a one-dimensional chain of length $L$.
The nearest-neighbor hopping amplitude is set to $t=1$ to fix the energy scale. 
Furthermore, $\mu$ denotes a local potential, which allows us to tune between topological distinct phases. 
The remaining terms in Eq.\ (\ref{H_Kit}) represent $p$-wave superconducting pairing with the complex gap parameter 
\be
\Delta = \Delta_{0} e^{i\phi} 
\: , 
\label{deltacom}
\ee
where $\Delta_{0}$ and $\phi$ denote the real amplitude and phase respectively.
Introducing Majorana operators, 
\ba
\gamma_{j}^{A} &=& \left( \gamma_{j}^{A} \right)^{\dagger} = c_{j}^{\dagger} + c_{j}  
\: , 
\nonumber \\
\gamma_{j}^{B} &=& \left( \gamma_{j}^{A} \right)^{\dagger} = i \left( c_{j}^{\dagger} - c_{j}\right)
\: , 
\label{maj_ops}
\ea
and assuming for the moment that $\phi=0$, 
the Hamiltonian (\ref{H_Kit}) can be rewritten as 
\ba
H_{\text{Kit}} 
&=&
\frac{i}{2} \, \sum_{j=1}^{L-1} \Big[ 
-\mu \gamma_{j}^{A}\gamma_{j}^{B} 
\nonumber \\
&+& (\Delta_{0}+t) \gamma_{j}^{B}\gamma_{j+1}^{A} + (\Delta_{0}-t) \gamma_{j}^{A}\gamma_{j+1}^{B}
\Big]
\: .
\label{kitaev_maj}
\ea
Adding a hopping $t$ and a gap parameter $\Delta_{0}$ to link the sites $j=L$ and $j=1$, i.e., using periodic boundary conditions, makes the system invariant under discrete lattice translations $R_{j}$. 
Via Fourier transformation in the form $\gamma_{j}^{A/B} = L^{-1/2}\sum_{k} e^{-ikR_{j}} \gamma^{A/B}_{k}$, we obtain the $k$-space representation of the Hamiltonian: 
\be
H_{\text{Kit}} 
= 
\sum_{k} \left(
\gamma_{-k}^{A}, \gamma_{-k}^{B}
\right) 
h_{\text{Kit}}(k) 
\begin{pmatrix} \gamma_{k}^{A} \\ \gamma_{k}^{B} \end{pmatrix}
\: , 
\ee
where 
\be
h_{\text{Kit}}(k) 
= 
- 2 \Delta_{0} \sin(k) \sigma_{x} + ( \mu + 2t \cos(k)) \sigma_{y} 
\: , 
\label{bdgmat}
\ee
with Pauli matrices $\sigma_{x}$ and $\sigma_{y}$, 
takes the form of a $2 \times 2$ Bogoliubov–de Gennes (BdG) matrix \cite{Chiu2016}.

The pairing term explicitly breaks the U(1) charge symmetry such that the total particle number $N=\sum_{j} c_{j}^{\dagger} c_{j}$ is not longer conserved, while fermion parity $(-1)^{N}$ is still a good quantum number.
Furthermore, the Kitaev-chain Hamiltonian, Eq.\ (\ref{H_Kit}), transforms as
\be
K H_{\text{Kit}} K^{\dagger} = -H_{\text{Kit}}^{\ast}
\label{H_phc}
\ee
under particle-hole conjugation $K$. Here, $H_{\text{Kit}}^{\ast}$ denotes the Hamiltonian, Eq.\ (\ref{H_Kit}), with complex conjugate coefficients. The unitary particle-hole conjugation operator $K$ is defined via its action 
\be
K c_{j} K^{\dagger} = c_{j}^{\dagger} 
\: , \quad 	
K c_{j}^{\dagger} K^{\dagger} = c_{j}
\: , 
\label{phc_kit}
\ee
on the elementary fermionic annihilation and creation operators $c_{j}$ and $c_{j}^{\dagger}$  \cite{Zirnbauer2021}.

The Kitaev-chain model features two topologically distinct phases, distinguished by a topological $\mathbb{Z}_{2}$ invariant 
\begin{align}
\nu = \lvert \text{deg}(m) \rvert \: ,
\label{nu}
\end{align}
which can be derived from the degree
\be
\text{deg}(m) = \frac{1}{2 \pi} \int \sum_{ab} \epsilon^{ab} m_{a}(k) \frac{\partial m_{b}(k)}{\partial k} \text{d} k
\ee
($a,b = 1,2$ and $\epsilon^{ab} = - \epsilon^{ba} = 1$ for $a \ne b$ and zero else)
of the map 
\ba
m : \mathbb{S}^{1}_{k} &\to& \mathbb{S}^{1} \subset \mathbb{R}^{2} 
\nonumber \\
k &\mapsto& \mathbf{m}(k) 
= 
\begin{pmatrix}
- 2 \Delta_{0} \sin(k) \\
\mu + 2t \cos(k)
\end{pmatrix} 
\: .
\label{eq:m}
\ea
We note that the map $m$ can be used to express the BdG matrix in Eq.\ (\ref{bdgmat}) as 
$h_{\text{Kit}}(k) = \mathbf{m}(k)^{\intercal} \cdot \ff \sigma$.
If the gap parameter is complex with $\phi \ne 0$, the Hamiltonian is given by the same form of the BdG matrix but with 
$\ff \sigma = (\sigma_{x}, \sigma_{y}, \sigma_{z})^{\intercal}$ and with a different map $\widetilde{m} : \mathbb{S}^{1}_{k} \to \mathbb{S}^{1} \subset \mathbb{R}^{3}$, where
\ba
\widetilde{\mathbf{m}}(k) 
= 
\begin{pmatrix}
- 2 \Delta_{0} \sin(k) \cos(\phi) \\
\mu + 2t \cos(k) \\
-2 \Delta_{0} \sin(k) \sin(\phi)
\end{pmatrix} 
\: .
\label{eq:m_tilde}
\ea
However, if Eq.\ (\ref{eq:m}) is interpreted as a map $m : \mathbb{S}^{1}_{k} \to \mathbb{S}^{1} \subset \mathbb{R}^{3}$, by adding a trivial third component, it can be continuously deformed into Eq.\ (\ref{eq:m_tilde}) for every value of the superconducting phase $\phi$. 
As a result, the degree, and thus the topological invariant $\nu$, associated with $\mathbf{m}(k)$ and $\widetilde{\mathbf{m}}(k) $ is the same.
Equation (\ref{nu}) yields two parameter regimes.
For $2 |t| > |\mu|$ we have $\nu = 1$, i.e., the system is in a topologically nontrivial state, while a trivial state with $\nu = 0$ is realized if $2 |t| < |\mu|$. 

\section{Theoretical concepts}
\label{sec:meth}

\subsection{Wilczek-Zee phase}
\label{sec:wz}

$H_{\text{Kit}}$, as given by Eq.\ (\ref{H_Kit}), can be understood as a continuous family of Hamiltonians $H_{\text{Kit}}(\phi)$ parameterized by the phase $\phi \in \mathbb{S}^{1}_{\phi}$ of the superconducting gap parameter $\Delta = \Delta_{0} e^{i\phi}$. 
This also generates a continuous family of many-body energy eigenstates $\ket{n(\phi)}$ over the parameter manifold $\mathbb{S}^{1}_{\phi}$, satisfying $H_{\text{Kit}}(\phi)  \ket{n(\phi)} = E_{n}(\phi) \ket{n(\phi)}$.
In the topologically nontrivial $\nu=1$ bulk phase, there are two degenerate ground states due to the presence of a pair of Majorana zero modes (MZMs) localized at the chain boundaries as enforced by the bulk-boundary correspondence. 
The ground-state degeneracy can be higher and more MZMs can be present when networks of Kitaev chains are considered. 
If there is a finite overlap of boundary modes, either due to a finite (or even small) chain length or due to weakly linked chains, the boundary-mode energies will split, and thus the ground-state degeneracy will be lifted.
Note that we will nevertheless continue to refer to those modes as MZMs for simplicity.

Rather than adiabatically exchanging MZMs in real space \cite{Mascot2023,Halperin2012,Kraus2013,Amorim2015,Sekania2017,Harper2019,Tutschku2020,Tanaka2022}
and tracing the resulting effect of the unitary rotation in the ground-state sector picked up during this process, we will here consider {\em exchangeless} braiding. 
Our approach is to consider the transport of MZMs along closed continuous paths in the parameter manifold $\mathbb{S}^{1}_{\phi}$, i.e., a continuous change of the phase of the superconducting order parameter by an integer multiple of $2\pi$. 
Generally, for topological defect modes in topological $p$-wave superconductors, we expect a non-Abelian Ising anyon statistics for this process. 

To demonstrate this, we consider the $d_{0}=2^{q}$-dimensional subspace $\mathcal{H}_{0}(\phi)$ spanned by the instantaneous zero-energy (or non-degenerate low-energy) many-body states $\ket{n_{0}(\phi)}$, where $n_{0} = 1, 2, ... , d_{0}$ and $q$ is the number of MZM pairs in the system. 
A closed path in the parameter manifold $\mathbb{S}^{1}_{\phi}$ is given by a continuous map
\be
C : [0,1] \to \mathbb{S}^{1}_{\phi} \: , \qquad t \mapsto \phi = C(t)
\ee
with $C(0)=C(1)$.
The mismatch between a set of orthonormal basis states $\ket{n_{0}(\phi)}$ of $\mathcal{H}_{0}(\phi)$ at $\phi = C(0)$ and, after a cyclic parameter change, at $\phi=C(1)$, is quantified by the non-Abelian Wilczek-Zee (WZ) phase matrix \cite{Wilczek1984,Nayak1996}
\be
U_{\text{WZ}}[C] = \mathcal{P} \exp{\left[i\oint_{C} d \phi \, A(\phi) \right] }
\: .
\label{WZ}
\ee
Here, $\mathcal{P}$ denotes the path-ordering operator and $A(\phi)$ is the matrix-valued (non-Abelian) Berry connection with elements
\be
A_{n_{0}n'_{0}}(\phi) = i \bra{n_{0}(\phi)} \frac{\partial}{\partial \phi} \ket{n'_{0}(\phi)} 
\: .
\label{bconn}
\ee

It is important to distinguish between two cases, see, e.g., Ref.\ \cite{Bohm2003}.
In case (i), the energies of the basis states $\ket{n_{0}(\phi)}$ are degenerate for all $\phi \in \mathbb{S}^{1}_{\phi}$, and thus the choice of the orthonormal basis states $\ket{n_{0}(\phi)}$ is arbitrary for each $\phi$. 
Observable quantities must be invariant under a local gauge transformation 
$\ket{n_{0}(\phi)} \mapsto \sum_{n'_{0}} G_{n'_{0}n_{0}}(\phi) \ket{n'_{0}(\phi)}$
given by an arbitrary unitary $d_{0}$-dimensional matrix $G(\phi) \in \mbox{U($d_{0}$})$, which smoothly depends on $\phi$. 
The Berry connection, Eq.\ (\ref{bconn}) transforms non-trivially under $G(\phi)$, i.e., $A(\phi) \mapsto G^{\dagger}(\phi) A(\phi) G(\phi) + i G^{\dagger}(\phi)\partial_{\phi} G(\phi)$.
However, the WZ phase matrix is gauge covariant, $U_{\text{WZ}}[C] \mapsto G^{\dagger}[C] \, U_{\text{WZ}}[C] \, G[C]$, such that $\mbox{tr} (U_{\text{WZ}}[C])$ and 
$\mbox{det} (U_{\text{WZ}}[C])$ are gauge-invariant observables.
Note that for a one-dimensional subspace ($d_{0}=1$), the WZ phase is just the Abelian and gauge invariant Berry phase \cite{Ber84}
$e^{i\gamma_{\rm B}} = \exp(i \oint_{C} d \phi \, A(\phi))$.

Here, we are concerned with case (ii), where the energies of the eigenstates $\ket{n_{0}(\phi)}$ with $n_{0}=1,...,d_{0}$ spanning the low-energy subspace $\mathcal{H}_{0}(\phi)$ with dimension $d_{0} > 1$ are non-degenerate, except for degeneracies at isolated points along $\mathbb{S}^{1}_{\phi}$.
In this case, there is much less gauge freedom, as the eigenstates are unique up to U(1) phase factors only, i.e., the gauge group reduces to $T = {\rm U}(1) \times \cdots \times {\rm U}(1) <  \mbox{U}(d_{0})$. 

We compute the WZ phase numerically using a discretized version of Eq.\ (\ref{WZ}), 
\be
U_{\text{WZ}}[C] \approx U_{\text{WZ}}[C_{I}] = \prod_{i=0}^{I-1} a(i)
\: , 
\label{WZ_disc}
\ee
where the elements of the of the $d_{0} \times d_{0}$-dimensional matrices $a(i)$ are given by
\be
a_{n_{0}n'_{0}}(i) = \braket{n_{0}(\phi_{i}) | n'_{0}(\phi_{i+1})}
\: , 
\label{a_aux}
\ee
i.e., by the overlaps between instantaneous low-energy eigenstates of $\mathcal{H}_{0}(\phi_{i})$ and of $\mathcal{H}_{0}(\phi_{i+1})$. 
Furthermore, $C_{I}$ in Eq.\ (\ref{WZ_disc}) is a discretization 
\be
C_{I} = \left\{\phi_{0},\phi_{1},...,\phi_{I-1},\phi_{I} = \phi_{0}\right\}
\label{C_I}
\ee
of the original continuous loop $C$ in $\mathbb{S}_{\phi}^{1}$. The parameter $I$ controls the resolution of the discretized loop.

\subsection{Bogoliubov diagonalization}

An evaluation of Eq.\ (\ref{WZ_disc}) requires a numerically stable way of calculating the overlaps, Eq.\ (\ref{a_aux}), between instantaneous many-body eigenstates of the Kitaev-chain Hamiltonian given by Eq.\ (\ref{H_Kit}).
To this end, we will use Bogoliubov-de Gennes theory, i.e., we rewrite the Hamiltonian in a BdG form as
\be
H_{\text{Kit}} = \frac{1}{2} \ff \Psi^{\dagger} h_{\text{Kit}} \ff \Psi
\label{kcbdg}
\; , 
\ee
where $\ff \Psi = \left(c_{1},...,c_{L},c_{1}^{\dagger},...,c_{L}^{\dagger}\right)^{\intercal}$ is the Nambu spinor built from the fermionic annihilators and creators,
and where
\be
h_{\text{Kit}} 
= 
\begin{pmatrix}
T & \Delta \\
\Delta^{\dagger} & -T^{\ast}
\end{pmatrix}
\label{h_Kit}
\ee
is the $2L \times 2L$ BdG matrix comprized of the $L \times L$ single-particle hopping matrix $T$ with elements
\be
T_{jj'} = -t (\delta_{j,j'+1}+\delta_{j+1,j'}) - \mu \delta_{jj'} 
\ee
and the $L \times L$ gap matrix $\Delta$ defined by its elements
\be
\Delta_{jj'} = \Delta (\delta_{j+1,j'} - \delta_{j,j'+1})
\: .
\ee

Since the Hamiltonian $H_{\text{Kit}}$ and thus $h_{\text{Kit}}$ are Hermitian, $T$ must be Hermitian and $\Delta$ antisymmetric.
The transformation of $H_{\text{Kit}}$ under particle-hole conjugation $K$, as given in Eq.\ (\ref{H_phc}), implies that the matrix $h_{\text{Kit}}$ transforms as
\be
\Sigma \, h_{\text{Kit}} \, \Sigma = - h^{\ast}_{\text{Kit}}
\ee
under the matrix representation
\be
\Sigma = \sigma_{x} \otimes \mathbb{1}_L 
\ee
of $K$ in the Nambu spinor basis. 
A unitary diagonalization of $h_{\text{Kit}}$ must respect this symmetry to ensure that the resulting quasi-particle annihilators and creators, $b_{m}$ and $b_{m}^{\dagger}$, satisfy fermionic anti-commutation relations.
In a first step, we therefore consider the transformation
\be
\bar{h}_{\text{Kit}} = W h_{\text{Kit}} W^{\dagger} 
\ee
with 
\be
W = \frac{1}{\sqrt{2}}
\begin{pmatrix}
1 & -i \\
i & -1
\end{pmatrix} 
\otimes \mathbb{1}_L
= W^{\dagger} = W^{-1}
\: .
\ee
This yields a real and skew-symmetric $2L \times 2L$ matrix $i \bar{h}_{\text{Kit}}$ with an eigenvalue spectrum $\{\pm i e_{1},...,\pm i e_{L}\}$ of pairwise conjugate imaginary eigenvalues. Using a numerical real Schur decomposition algorithm, $i \bar{h}_{\text{Kit}}$ can be brought into the form
\ba
	\bar{E} = \begin{pmatrix}
		\mathbb{0}_{L} & E \\
		-E & \mathbb{0}_{L}
	\end{pmatrix} = (i\sigma_{y}) \otimes E
\ea
by a $2L$-dimensional real and orthogonal matrix $O$, i.e. 
\be
	O^{\intercal} (i\bar{h}_{\text{Kit}}) O = \bar{E}.
	\label{OhO}
\ee
Here, $E = \text{diag}(E_{1},...,E_{L})$ is the $L \times L$ diagonal matrix with $E_{m} = |i e_{m}|$, and $\mathbb{0}_{L}$ is the $L \times L$ zero matrix.
One easily verifies that $B = W^{\dagger} O W$ is a Bogoliubov matrix of the form
\be
B = 
\begin{pmatrix}
U & V^{\ast} \\
V & U^{\ast} 
\end{pmatrix}
\ee
with $\Sigma B \Sigma =  B^{\ast}$ and $E_{\text{Kit}} = - i W^{\dagger} \bar{E} W$ is a real diagonal-matrix of the form
\be
E_{\text{Kit}} = 
\begin{pmatrix}
	E & \mathbb{0}_{L} \\
	\mathbb{0}_{L} & -E
\end{pmatrix} = \sigma_{z} \otimes E,
\ee
such that Eq.\ (\ref{OhO}) immediately yields a Bogoliubov diagonalization 
\be
B^{\dagger} h_{\text{Kit}} B = E_{\text{Kit}}.
\label{BhB}
\ee
of the Kitaev chain BdG matrix Eq.\ (\ref{h_Kit}).

With the matrices $U$ and $V$ at hand, we can define a new Nambu spinor
\be
\ff \Phi
= 
\begin{pmatrix}
\ff b \\
\ff b^{\dagger}
\end{pmatrix} 
= 
\begin{pmatrix}
U^{\dagger} & V^{\dagger} \\
V^{\intercal} & U^{\intercal}
\end{pmatrix} 
\begin{pmatrix}
\ff c \\
\ff c^{\dagger}
\end{pmatrix} 
=   
B^{\dagger} \,  \ff \Psi
\: 
\label{Bogolions_1} 
\ee
and thus the annihilators and creators $b_{m}$ and $b_{m}^{\dagger}$ of fermionic Bogoliubov quasi-particles.
Furthermore, $B$ diagonalizes the Hamiltonian $H_{\text{Kit}}$ in Eq.\ (\ref{kcbdg}), 
\ba
H_{\text{Kit}} 
&=& 
\frac{1}{2} \ff \Psi^{\dagger} h_{\text{Kit}} \ff \Psi
= 
\frac{1}{2} \ff \Psi^{\dagger} B \, E_{\text{Kit}} \,  B^{\dagger} \ff \Psi
\nonumber \\
&=& 
\frac{1}{2} \ff \Phi^{\dagger} E_{\text{Kit}} \ff \Phi
= 
\sum_{m=1}^{L} E_{m} b_{m}^{\dagger} b_{m}
\; .
\label{H_Kit_BdG}
\ea

\subsection{Construction of ground states}

Still, the construction of a ground state as a quasi-particle vacuum is non-trivial. 
Generally, a non-trivial quasiparticle vacuum can be expressed in a number of different ways, depending on the invertibility of the Bogoliubov matrices $U$ and $V$.
When $V$ is invertible (see also the discussion further below), $\ket{0}_{b}$ may be written as a product state, indicated by ``p'' \cite{Shi2017,Mascot2023}
\be
\ket{0}_{b}^{\text{p}} 
= 
\prod_{m=1}^{L} b_{m} \ket{0}
\: , 
\label{0_p}
\ee
where $\ket{0}$ is the vacuum of the original fermions, i.e., $c_{j} \ket{0} = 0$ for all $j$.
With Eq.\ (\ref{H_Kit_BdG}), we immediately have $H_{\rm Kit} \ket{0}_{b}^{\text{p}} = E_{\rm GS} \ket{0}_{b}^{\text{p}}$ with $E_{\rm GS}=0$.
The norm of the product state $\ket{0}_{b}^{p}$ is given by 
\be
\mathcal{N}_{b}^{p} = (-1)^{\frac{L(L-1)}{2}}\text{Pf} 
\begin{pmatrix}
 V^{\intercal} U  & V V^{\ast} \\
-V^{\dagger} V  & U^{\dagger} V^{\ast} 
\end{pmatrix}
\label{sqrtnp}
\: , 
\ee
where $\text{Pf}$ denotes the Pfaffian.  
Physically, the invertibility of $V$ ensures that there is no Bogoliubov mode that annihilates the vacuum $\ket{0}$ of the original fermions, so that the construction remains non-trivial.

When $U$ is invertible, $\ket{0}_{b}$ may be written in the form of a so-called Thouless state \cite{Shi2017,Robledo2011}
\be
\ket{0}_{b}^{\text{T}} 
= 
\exp\left(\frac{1}{2} \ff c^{\dagger} \, S \, \left(\ff c^{\dagger}\right)^{\intercal}\right) \ket{0}
\: , 
\label{0_T}
\ee
where $S=\left(VU^{-1}\right)^{\ast}$, and where
\be 
\mathcal{N}_{b}^{T} 
= 
(-1)^{\frac{L(L+1)}{2}}
\text{Pf} 
\begin{pmatrix}
S  & -\mathds{1}_{L} \\
\mathds{1}_{L} & -S^{\ast} 
\end{pmatrix}
\label{sqrtnt}
\ee
is the norm of $\ket{0}_{b}^{T}$. 
When both $\ket{0}_{b}^{p}$ and $\ket{0}_{b}^{T}$ exist, they are related via \cite{Shi2017}
\be
\ket{0}_{b}^{\text{p}} = \text{Pf}\left(U^{\dagger}V^{\ast}\right) \ket{0}_{b}^{\text{T}}
\: .
\ee
There is a criterion for the invertibility of $U$ and $V$: 
When $L$ is odd, $\det(B)=-1$ prevents $U$ from being invertible, while $\det(B)=1$ prevents $V$ from being invertible. 
If $L$ is even and $\det(B)=-1$, neither $U$ nor $V$ can be inverted. 
In addition, it is possible that $U=U(\phi)$ or $V=V(\phi)$ are accidentally singular at isolated points $\phi \in \mathbb{S}^{1}_{\phi}$.

\subsection{Bloch-Messiah decomposition}

Here, our strategy is to construct the quasiparticle vacuum in the product form Eq.\ (\ref{0_p}), regardless of whether $V$ is singular or not.
This can be achieved with the Bloch-Messiah decomposition (BMD) \cite{Bloch1962}.
The BMD of a given $2L \times 2L$ Bogoliubov matrix $B$ is a factorization 
\be
B 
= 
\begin{pmatrix}
U & V^{\ast} \\
V & U^{\ast}
\end{pmatrix} 
= 
\begin{pmatrix}
C^{\dagger} & 0 \\
0 & C^{\intercal}
\end{pmatrix} 
\begin{pmatrix}
\bar{U} & \bar{V} \\
\bar{V} & \bar{U}
\end{pmatrix} \begin{pmatrix}
D & 0 \\
0 & D^{\ast}
\end{pmatrix}
= 
\mathcal{C}  \bar{B}  \mathcal{D}^{\dagger} 
\label{BMD}
\ee
of $B$ into a product of block diagonal unitary matrices $\mathcal{C}$ and $\mathcal{D}^{\dagger}$, which are defined in terms of unitary $L \times L$ matrices $C$ and $D$, and a real matrix $\bar{B}$. 
The $L \times L$ blocks of $\bar{B}$ are given by the diagonal matrix 
\be
\bar{U} 
= 
\begin{pmatrix}
\mathbb{0}_{\rm F}  &&  \\
& \bigoplus_{p =1}^{\rm P} u_{\mathrlap{\phantom{\bar{p}}}p} \mathds{1}_{2} 	& \\
&  &  & \mathds{1}_{\rm E} \\
\end{pmatrix} 
\label{Ubar_}
\ee
and by the block-diagonal matrix 
\be
\bar{V} 
= 
\begin{pmatrix}
\mathds{1}_{\rm F}  &&  \\
& \bigoplus_{p = 1}^{\rm P} i v_{\mathrlap{\phantom{\bar{p}}}p} \sigma_{y} 	& \\
&  &  & \mathbb{0}_{\rm E} \\
\end{pmatrix}
\: , 
\label{Vbar_}
\ee
where $u_{\mathrlap{\phantom{\bar{p}}}p}^{2} + v_{\mathrlap{\phantom{\bar{p}}}p}^{2} = 1$ and $u_{\mathrlap{\phantom{\bar{p}}}p}, v_{\mathrlap{\phantom{\bar{p}}}p}> 0$ for all $p = 1,...,d_{\rm P}$. 
Furthermore, 
$\mathbb{0}_{\rm F}$ and $\mathbb{0}_{\rm E}$ denote the $d_{\rm F}$- and the $d_{\rm E}$-dimensional zero matrix, and 
$\mathds{1}_{\rm F}$ and $\mathds{1}_{\rm E}$ the $d_{\rm F}$- and the $d_{\rm E}$-dimensional unit matrix.

Technically, for a given Bogoliubov matrix $B$, the numerical implementation of the BMD approach is straightforward and provides us with the matrices $C$, $D$ and $\bar{U}$, $\bar{V}$ in Eq.\ (\ref{BMD}).
Hence, we may consider these matrices as given. Inserting the BMD of $B$ into Eq.\ (\ref{Bogolions_1}), one finds
\be
\begin{pmatrix}
D & 0 \\
0 & D^{\ast}
\end{pmatrix}  
\begin{pmatrix}
\ff b \\ \ff b^{\dagger}
\end{pmatrix}
= 
\begin{pmatrix}
\bar{U} & \bar{V}^{\intercal} \\
\bar{V}^{\intercal} & \bar{U}
\end{pmatrix} 
\begin{pmatrix}
C & 0 \\
0 & C^{\ast}
\end{pmatrix}  
\begin{pmatrix}
\ff c \\ \ff c^{\dagger}
\end{pmatrix}
\: . 
\label{Bogolions_2}
\ee
This, and the unitarity of $C$ and $D$ motivates the definition
\be
\bar{c}_{r} 
= 
\sum_{j=1}^{L} C_{rj} c_{j} \; , \qquad \bar{b}_{n} = \sum_{m=1}^{L} D_{nm} b_{m}
\: .
\label{c_phi_bar}
\ee
With Eqs.\ (\ref{Bogolions_2}) and (\ref{c_phi_bar}) we also have
\be
\begin{pmatrix}
\bar{\ff b} \\
\bar{\ff b}^{\dagger}
\end{pmatrix}
= 
\begin{pmatrix}
\bar{U} & \bar{V}^{\intercal} \\
\bar{V}^{\intercal} & \bar{U}
\end{pmatrix} 
\begin{pmatrix}
\bar{\ff c} \\
\bar{\ff c}^{\dagger}
\end{pmatrix}
\label{Bogolions_3}
\: .
\ee
The fermionic annihilators $\bar{c}_{r}$ refer to one-particle states forming the so-called ``canonical'' single-particle basis, in which the one-particle reduced density matrix $\rho$ is diagonal \cite{Ring1980}.
According to Eq.\ (\ref{c_phi_bar}), annihilators of the original fermions are transformed among themselves, and analogously for annihilators of BdG quasiparticles.
This immediately implies that they belong to the same fermion vacuum and quasi-particle vacuum respectively.

The (block-)diagonal structures Eq.\ (\ref{Ubar_}) and Eq.\ (\ref{Vbar_}) of $\bar{U}$ and $\bar{V}$ allow us to identify three types of quasi-particles: 
(i) the $f=1,...,d_{\rm F}$ ``filled'' modes, where $v_{f} = 1$ and $u_{f} = 0$ such that $\bar{b}_{f} = \bar{c}_{f}^{\dagger}$; 
(ii) the $e=1,...,d_{\rm E}$ ``empty'' modes, where $v_{e} = 0$ and $u_{e} = 1$ such that $\bar{b}_{e} = \bar{c}_{e}$; 
and (iii) the paired modes, where $v_{\mathrlap{\phantom{\bar{p}}}p} , u_{\mathrlap{\phantom{\bar{p}}}p} > 0$ such that 
\ba
\bar{b}_{\mathrlap{\phantom{\bar{p}}}p} 
&=& 
u_{\mathrlap{\phantom{\bar{p}}}p} \bar{c}^{\phantom{\dagger}}_{\mathrlap{\phantom{\bar{p}}}p} 
-
v_{\mathrlap{\phantom{\bar{p}}}p} \bar{c}_{\bar{p}}^{\dagger} 
\: ,
\nonumber \\
\bar{b}_{\bar{p}} 
&=& 
u_{\mathrlap{\phantom{\bar{p}}}p} \bar{c}^{\phantom{\dagger}}_{\bar{p}} 
+ 
v_{\mathrlap{\phantom{\bar{p}}}p} \bar{c}_{\mathrlap{\phantom{\bar{p}}}p}^{\dagger}
\: . 
\ea
Here, $p,\bar{p}$ label the two entries of the $p$-th $2 \times 2$ block on the diagonal of $\bar{V}$ and $\bar{U}$. 
We have $d_{\rm F} + 2 d_{\rm P} + d_{\rm E} = L$ by construction.

Since the $\ff b$ and $\bar{\ff b}$ quasi-particle annihilators share the same vacuum, we can simply rewrite Eq.\ (\ref{0_p}) in terms of the $\bar{\ff b}$ operators as 
\be
\ket{0}_{b}^{\text{p}} 
= 
\prod_{n=1}^{L} \bar{b}_{n} \ket{0}
\: .
\label{ref}
\ee
However, if there is at least one empty mode with $\bar{b}_{e} = \bar{c}_{e}$ ($d_{\rm E}>0$), 
it annihilates the original fermion-particle vacuum $\ket{0}$, since according to Eq.\ (\ref{c_phi_bar}),
$\bar{c}_{j}  \ket{0} = 0$.
In this case, the product state becomes trivial, $\ket{0}_{b}^{\text{p}} = 0$, and thus the construction of a ground state fails. 
With Eq.\ (\ref{Vbar_}) we see that $d_{\rm E}>0$ means that $\bar{V}$ is singular. 
But $\bar{V}$ is singular if and only if the original BdG matrix $V$ is singular, since the BMD Eq.\ (\ref{BMD}) yields 
$V = C^{\intercal} \bar{V} D^{\ast}$ with unitary $C$ and $D$.
This makes it quite clear why the product form of the quasi-particle vacuum is ill-defined when $V$ is singular.

However, the BMD allows us to identify and to truncate the empty modes in the construction of the ground state. 
In fact, we can simply define
\ba
\ket{\bar{0}}_{b}^{\rm p} 
&=& 
\frac{1}{\mathcal{N}_{\text{p}}} \prod_{p = 1}^{d_{\rm P}} \bar{b}_{\mathrlap{\phantom{\bar{p}}}p} \bar{b}_{\bar{p}} 
\prod_{f =1}^{d_{\rm F}} \bar{b}_{f} \ket{0} 
\nonumber \\
&=& 
\prod_{p = 1}^{d_{\rm P}} 
\left( u_{\mathrlap{\phantom{\bar{p}}}p} + v_{\mathrlap{\phantom{\bar{p}}}p} 
\bar{c}_{\mathrlap{\phantom{\bar{p}}}p}^{\dagger} \bar{c}_{\bar{p}}^{\dagger}\right) 
\prod_{f =1}^{d_{\rm F}} \bar{c}_{f}^{\dagger} \ket{0}
\: .
\label{Bogo_vac_trun}
\ea
Regardless of whether $V$ is singular ($d_{\rm E}>0$) or not ($d_{\rm E}=0$), the quasi-particle vacuum $\ket{\bar{0}}_{b}^{\rm  p}$ is non-trivial and yields a ground state of $H_{\rm Kit}$.
Note that Eq.\ (\ref{Bogo_vac_trun}) immediately implies that the total number of Bogoliubov quasi-particles of $\ket{\bar{0}}_{b}^{\rm p}$ vanishes, while the total fermion particle number $N=\sum_{j=1}^{L} c_{j}^{\dagger} c_{j}$ depends on the number of filled modes $d_{\rm F}$.
	
\subsection{Overlap between excited states}

Based on the product form of the quasi-particle vacuum, Eq.\ (\ref{Bogo_vac_trun}), we can construct all excited energy eigenstates as BdG Fock states:
\be
\ket{n}_{b} 
= 
\ket{n_{1},...,n_{L}} 
= 
\prod_{m=1}^{L} (b_{m}^{\dagger})^{n_{m}} \, \ket{\bar{0}}_{b}^{p}
\: .
\label{Bogo_Fock}
\ee
Here, $n_{m} = 0,1$ denotes the occupation of the $m$-th quasi-particle state.
For the evaluation of the WZ phase, see Eqs.\ (\ref{WZ_disc}) and (\ref{a_aux}), we have to compute the overlap $\braket{n(\phi_{i}) | n'(\phi_{i+1})}$ of two arbitrary states of the form given by Eq.\ (\ref{Bogo_Fock}).

Many overlap formulas between BdG Fock states involve a square root.
Such formulas exhibit a sign ambiguity that, in reference to a paper by Onishi and Yoshida \cite{Onishi1966}, is known as the ``Onishi sign problem.''
To avoid this sign ambiguity, we adopt a Pfaffian formula that is due to Bertsch and Robledo \cite{Robledo1994,Robledo2009,Robledo2011,Bertsch2012}:
\be
\tensor[_b]{\braket{n_{r}^{\phantom{\prime}}|n_{s}^{\prime}}}{_b} = \pm \frac{1}{\mathcal{N}_{\text{p}}\mathcal{N}_{\text{p}}^{\prime}} \text{Pf}\left(\mathbb{M}\right)
\: .
\label{overlap}
\ee
Here
\be
\ket{n_{r}}_{b} 
= 
b_{m_{1}}^{\dagger} \cdots b_{m_{R}}^{\dagger} \ket{\bar{0}}_{b}^{p}
\ee
represents a BdG Fock state with $R$ excitations as specified by the ordered index set $r = \{m_{1},...,m_{R}\} \subset [1,L]$, 
and 
\be
\mathbb{M} 
= 
\begin{pmatrix}
\bar{V}^{\intercal}\bar{U} & \bar{V}^{\intercal}C^{\dagger}V^{\ast}_{r} & \bar{V}^{\intercal}C^{\dagger}U^{\prime}_{s} & \bar{V}^{\intercal}C^{\dagger}C^{\prime}\bar{V}^{\prime } \\
\cdot & \left(U_{r}\right)^{\dagger}V^{\ast}_{r} & \left(U_{r}\right)^{\dagger}U^{\prime}_{s} & \left(U_{r}\right)^{\dagger} C^{\prime}\bar{V}^{\prime } \\
\cdot &\cdot & \left(V^{\prime}_{s}\right)^{\intercal}U^{\prime}_{s} & \left(V^{\prime}_{s}\right)^{\intercal} C^{\prime}\bar{V}^{\prime } \\
\cdot &\cdot &\cdot & \bar{U}^{\prime} C^{\prime}\bar{V}^{\prime } 
\end{pmatrix}
\: , 
\ee
is a skew-symmetric matrix, where the $\bar{U}$, $\bar{V}$ and $C$ matrices belong to the BMDs of the Bogoliubov matrices $(U,V)$ that are associated to $\ket{n_{r}}_{b}$, see Eq.\ (\ref{BMD}), and analogously for $\bar{U}'$, $\bar{V}'$, and $C'$ associated to $\ket{n'_{s}}_{b}$. 
The $r$ and $s$ matrix subscripts indicate a restriction of the column index set onto the respective index set, $r$ or $s$. 
For instance, $U_{r}$ is an $L \times R$ matrix formed by the columns $m_{1}, ..., m_{R}$ of $U$. 
The sign $\eta=\pm 1$ in Eq.\ (\ref{overlap}) is given by 
\be
\eta = (-1)^{\frac{O(O-1)+R(R-1)}{2}}
\: , 
\ee
where $O = d_{\rm F} + 2 d_{\rm P}$ is the number of non-empty BdG modes in the vacuum $\ket{0}_{b}^{\text{p}}$ to which $\ket{n_{r}}_{b}$ adds $R$ excitations. 
The normalization factors in Eq.\ (\ref{overlap}) are $\mathcal{N}_{\text{p}} = \prod_{p=1}^{d_{\rm P}} v_{\mathrlap{\phantom{\bar{p}}}p}$ and $\mathcal{N}_{\text{p}}^{\prime} = \prod_{p'=1'}^{d_{\rm P'}} v_{p'}$, respectively.

Overlap formulas that rely on computing overlap squares by determinants require taking the square root to get the overlap. 
The Bertsch-Robledo formula in Eq.\ (\ref{overlap}) solves this problem by computing the overlap directly using the Pfaffian of a matrix instead. 
This removes the Onishi sign ambiguity.
There are several ways to derive Eq.\ (\ref{overlap}). 
The one used by Robledo is based on the Berezin integration of Grassmann numbers arising in the context of fermionic coherent states \cite{Robledo1994,Robledo2009,Robledo2011,Bertsch2012}. 
Another elegant way to arrive at Eq.\ (\ref{overlap}) was demonstrated by Neergard, who gave a combinatorial proof based on the interpretation of the Fock space representation as a spin representation \cite{Neergard2023}. 
In this interpretation, the Onishi sign ambiguity is a direct consequence of the double-valuedness of the spin representation of the orthogonal group.

\section{Exchangeless braiding in a single Kitaev chain}
\label{sec:one_KC}

Our goal is to apply the comprehensive many-body framework, as outlined in Sec.\ \ref{sec:meth} to demonstrate non-Abelian exchangeless braiding in networks of finite Kitaev chains. 
Specifically, we aim to showcase the persistence and robustness of simple topological quantum gates for cases, where the chains are so short or where the coupling between the chains is so strong that the energies of the MZMs significantly differ from zero.

As a starting point, we consider a Kitaev model on a single chain of length $L$ as described by Eq.\ (\ref{H_Kit}). 
In the topologically non-trivial phase and for $L\to \infty$, there are two MZMs $\gamma_{L}$ and $\gamma_{R}$ localized on the left and right boundary sites of the chain, respectively. 
The MZMs combine into a single complex zero-energy Bogoliubov mode with annihilation operator
\be
b_{0} = \frac{1}{2} \left(\gamma_{L} + i \gamma_{R} \right) 
\: . 
\label{complex_d0}
\ee
This allows us to construct two orthogonal many-body ground states
\be
\ket{0}_{b} \quad \text{and} \quad \ket{1}_{b} = b_{0}^{\dagger} \ket{0}_{b}
\: , 
\label{gs_1}
\ee
where we have written $\ket{0}_{b} \equiv \ket{\bar{0}}_{b}^{p}$ for simplicity, i.e., $\ket{0}_{b}$ denotes the truncated product form of the quasi-particle vacuum defined in Eq.\ (\ref{Bogo_vac_trun}). 
Both $\ket{0}_{b}$ and $\ket{1}_{b}$ are eigenstates of the Bogoliubov quasiparticle parity operator $P_{b} = (-1)^{N_{b}}$, where $N_{b} = \sum_{m} b_{m}^{\dagger}b_{m}$ is the total quasi-particle number operator. 
The ground states $\ket{0}_{b}$ and $\ket{1}_{b}$ have different parity. 
Their span defines the subspace $\mathcal{H}_{0}$.

In any chain of finite length $L < \infty$, the two boundary MZMs $\gamma_{L}$ and $\gamma_{R}$ are weakly coupled through an effective hybridization or ``interaction''. 
This effect can be described by a simple two-mode theory which has been proposed in Ref.\ \cite{Kitaev2001}.
The two-mode theory provides us with an effective Hamiltonian given by
\be
H_{\text{2-mode}}  
= 
\frac{i}{2} h \, \gamma_{L}\gamma_{R}
\: .
\label{eq:Maj_int}
\ee
Physically, the coupling strength $h$ corresponds to the amplitude for quasi-particle tunnelling from one end of the chain to the other.
For large $L$, it exponentially decays with $L$
\be
h \propto  e^{-L/\lambda}
\: .
\label{eq:Maj_int_h}
\ee
The characteristic length scale 
\be
\lambda = \big[\min \{ \vert \ln (\vert x_{+}\vert )\vert ,\vert \ln(\vert x_{-}\vert )\vert \}\big]^{-1}
\label{eq:int_length}
\: , 
\ee
can be expressed in terms of the so-called decay parameters
\be
x_{\pm} 
= 
\frac{-\mu \pm \sqrt{\mu^{2}-4(t^{2}-\Delta_{0}^{2})}}{2(t + \Delta_{0})}
\label{eq:x_pm}
\: ,  
\ee
as described in detail in Ref.\ \cite{Kitaev2001}.
The interaction $h$ in Eq.\ (\ref{eq:Maj_int}) removes the degeneracy of the ground-state energy and leads to an energy splitting of approximately $h$, i.e., a splitting that is exponentially small in $L$.
For simplicity, we will nevertheless continue to refer to $\gamma_{L}$ and $\gamma_{R}$ as MZMs.
In the full many-body theory, the low-energy eigenstates are still given by $\ket{0}_{b}$ and $\ket{1}_{b}$, rather than by a linear combination, since they are simultaneous eigenstates of $P_{b}$. 
{\em A priori}, however, it is not easy to find out, which of the two states is the actual ground state. 
Fortunately, this is not relevant for the rest of our considerations.

For $L\to \infty$, or if the interaction between the boundary MZMs $\gamma_{L}$ and $\gamma_{R}$ is negligibly small, the MZMs are projectively equivalent to Ising anyons \cite{Wilczek1998,Freedman2011}. 
This means that an adiabatic exchange of MZMs {\em in real space} would induce a unitary transformation $U_{LR}$ on the subspace $\mathcal{H}_{0}$ of degenerate many-body ground states that matches the Ising anyon braiding matrix \cite{Rowell2018}  
\be
R_{\text{Ising}} 
= 
e^{-i\pi/8} 
\begin{pmatrix}
	1 & 0 \\
	0 & e^{i\pi/2}
\end{pmatrix}
\: ,
\label{Ising_braiding}
\ee
up to a global phase factor, i.e., 
\be
U_{LR} = e^{i\alpha} R_{\text{Ising}}
\: .
\ee

Here, we consider exchangeless braiding \cite{Teo2010,Kitaev2001}. 
With a complex superconducting (SC) gap parameter $\Delta = \Delta_{0} e^{i\phi}$, the Kitaev-chain Hamiltonian in Eq.\ (\ref{H_Kit}) defines a continuous family of Hamiltonians $H_{\text{Kit}}(\phi)$ parameterized by the phase $\phi \in \mathbb{S}^{1}_{\phi}$ of $\Delta$.
For given $\phi$, the instantaneous many-body ground states $\ket{0(\phi)}_{b}$ and $\ket{1(\phi)}_{b}$ span the $\phi$-dependent subspace $\ca H_{0}(\phi)$.
Their instantaneous unitary evolution generated by $H_{\text{Kit}}(\phi)$ along closed curves $C$ in $\mathbb{S}^{1}_{\phi}$ is described 
by the WZ phase matrix $U_{\text{WZ}}[C]$, see Eq.\ (\ref{WZ}).
Below, we explicitly determine $U_{\text{WZ}}[C]$ for closed curves 
\ba
C^{N} : [0,1] &\to& \mathbb{S}^{1}_{\phi} 
\: ,
\nonumber \\
t &\mapsto & 2\pi N t
\: ,
\label{CN}
\ea
which cover the $\mathbb{S}^{1}_{\phi}$ parameter manifold $N$ times. 
$U_{\text{WZ}}[C]$ is calculated numerically via a discretization of $C^{N}$ with $I$ interpolation points, as described in Sec.\ \ref{sec:wz}; see Eqs.\ (\ref{WZ_disc}), (\ref{a_aux}), and (\ref{C_I}). 
The error is easily controlled by increasing the resolution $I\to \infty$.
In particular, the overlap matrices between states at consecutive points $\phi_{i}$, $\phi_{i+1}$ on the discretized curve $C_{I}^{N}$ witth $i=0,1,...,I-1$,
\be
a(i) = \begin{pmatrix}
\tensor[_b]{\braket{0(\phi_{i})|0(\phi_{j+1})}}{_b} & \tensor[_b]{\braket{0(\phi_{i})|1(\phi_{j+1})}}{_b} \\
\tensor[_b]{\braket{1(\phi_{i})|0(\phi_{i+1})}}{_b} & \tensor[_b]{\braket{1(\phi_{i})|1(\phi_{i+1})}}{_b}
\end{pmatrix} 
\label{a_mat}
\: , 
\ee
are evaluated using Eq.\ (\ref{overlap}). 

Let us first consider a Kitaev model with parameters $\Delta_{0}=t=1$ and $\mu = 0$ on a finite chain with $L=20$ sites. 
In this case, the MZMs do not hybridize, and the two states $\ket{0}_{b}$ and $\ket{1}_{b}$ have degenerate energies even for finite $L$.
The parameter set thus describes an ``ideal'' situation, which is reproduced by the two-mode theory, because the decay parameters $x_{\pm}$ in Eq.\ (\ref{eq:x_pm}) vanish, and which is perfectly suited to check the numerical implementation: 

Choosing the closed curve $C^{1}$ with winding $N=1$ and a resolution $I=10^{3}$, turns out as sufficient to obtain a numerical result for the WZ phase matrix with an absolute error for the matrix elements of less than $1\%$. 
The error is easily controlled and can be further suppressed by increasing the resolution. 
For the ideal case with $\Delta_{0}=t=1$ and $\mu=0$, we obtain the WZ phase matrix 
\be
U_{\text{WZ}}\left[C^{1}\right] 
= 
e^{i\alpha} \begin{pmatrix}
1 & 0 \\
0 & -1
\end{pmatrix}
\: . 
\label{WZ_C_1}
\ee
We will show in the following, that this result for $U_{\text{WZ}}\left[C^{1}\right]$ is rather generic and robust against perturbations lifting the ground-state degeneracy.

\begin{figure}[t] 
\centering
\includegraphics[width = 0.99\linewidth]{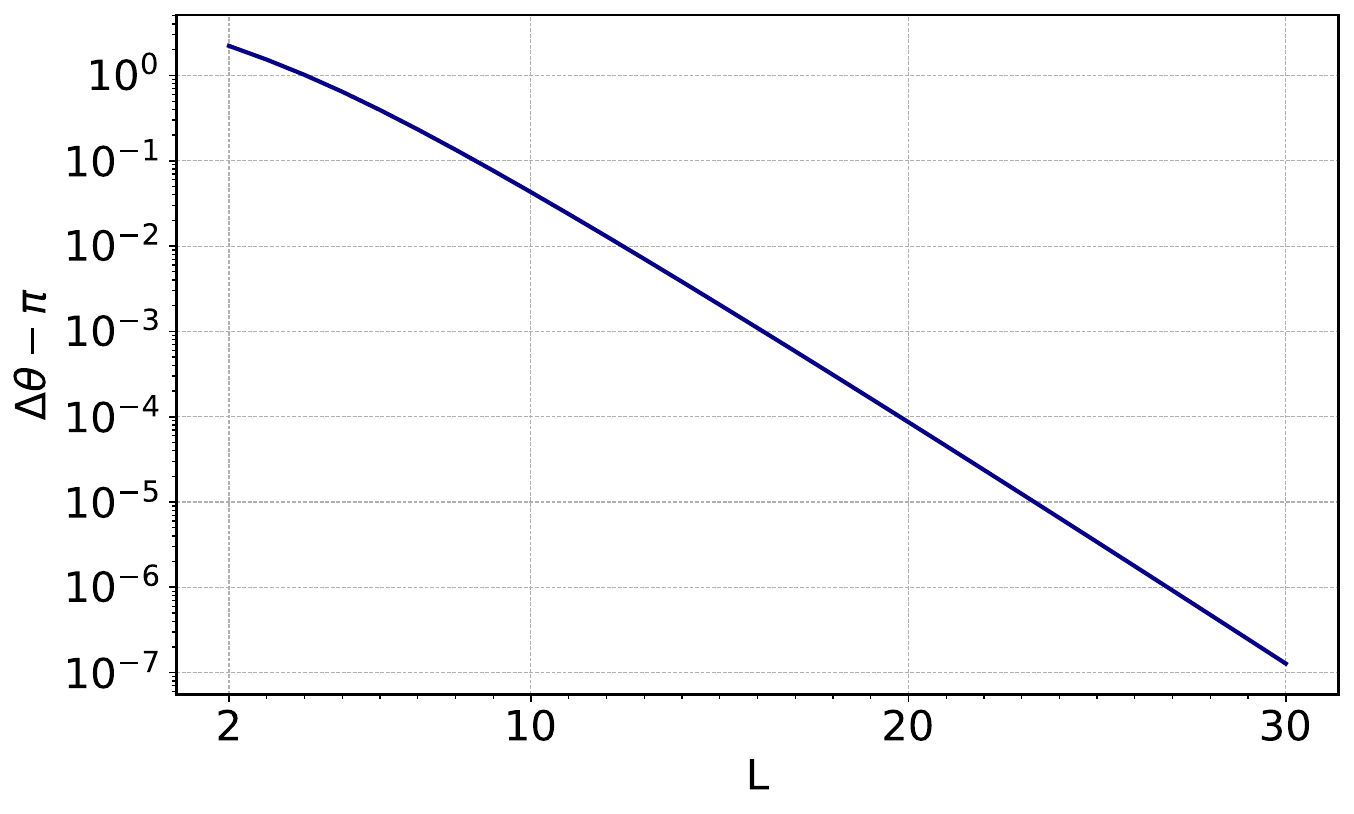}
\caption{
Deviation of the phase difference $\Delta\theta$ from $\pi$, see Eq.\ (\ref{D_theta}), as a function of $L$ for a single Kitaev chain. 
Parameters: $t = \Delta_{0}$ and $\mu=1$. 
Note the logarithmic scale.
}
\label{fig:delta_Z}
\end{figure}

Note that the WZ phase matrix given in Eq.\ (\ref{WZ_C_1}) is gauge invariant under arbitrary U(1) phase transformations of the two basis states $\ket{0}_{b} \equiv \ket{0(\phi_{0})}_{b}$ and $\ket{1}_{b} \equiv \ket{1(\phi_{0})}_{b}$, i.e., invariant under transformations in $T=\text{U}(1) \times \text{U}(1) < \text{U}(2)$, as discussed in Sec.\ \ref{sec:wz}.
It is {\em not} invariant under arbitrary $\text{U}(2)$ transformations, which 
form the larger gauge group that only applies to the case of a strictly two-fold degenerate ground-state energy, e.g., to the ``ideal'' parameter set or to the $L\to \infty$ limit.

Equation (\ref{WZ_C_1}) already represents a nontrivial result, because this WZ phase matrix is the generic form for a single Kitaev chain.
It corresponds to the square of the Ising anyon braiding matrix Eq.\ (\ref{Ising_braiding}). 
This shows that a $2\pi$ rotation of the phase of the SC gap parameter implements an exchangeless equivalent of a {\em double} exchange of Ising anyons in real space. 
This is what we call exchangeless braiding. 

The global U(1) phase factor $e^{i\alpha}$ in Eq.\ (\ref{WZ_C_1}) turns out as strongly parameter dependent (see below) but is physically insignificant. 
The Ising anyon statistics is rather captured by the 
gauge-invariant {\em difference}
\be
\Delta\theta\big(U_{\text{WZ}}[C^{1}] \big) 
= 
(\theta_{0} - \theta_{1}) \, \text{mod} \, 2 \pi = \pi
\, 
\label{D_theta}
\ee
between the phases $\theta_{0}$ and $\theta_{1}$ of $\ket{0}_{b}$ and $\ket{1}_{b}$, respectively. 
Since $\ket{0}_{b}$ and $\ket{1}_{b}$ have opposite parities, one may interprete $\Delta \theta$ as a U(1) Berry-phase difference.

\begin{figure}[t] 
\centering
\includegraphics[width = 1.0\linewidth]{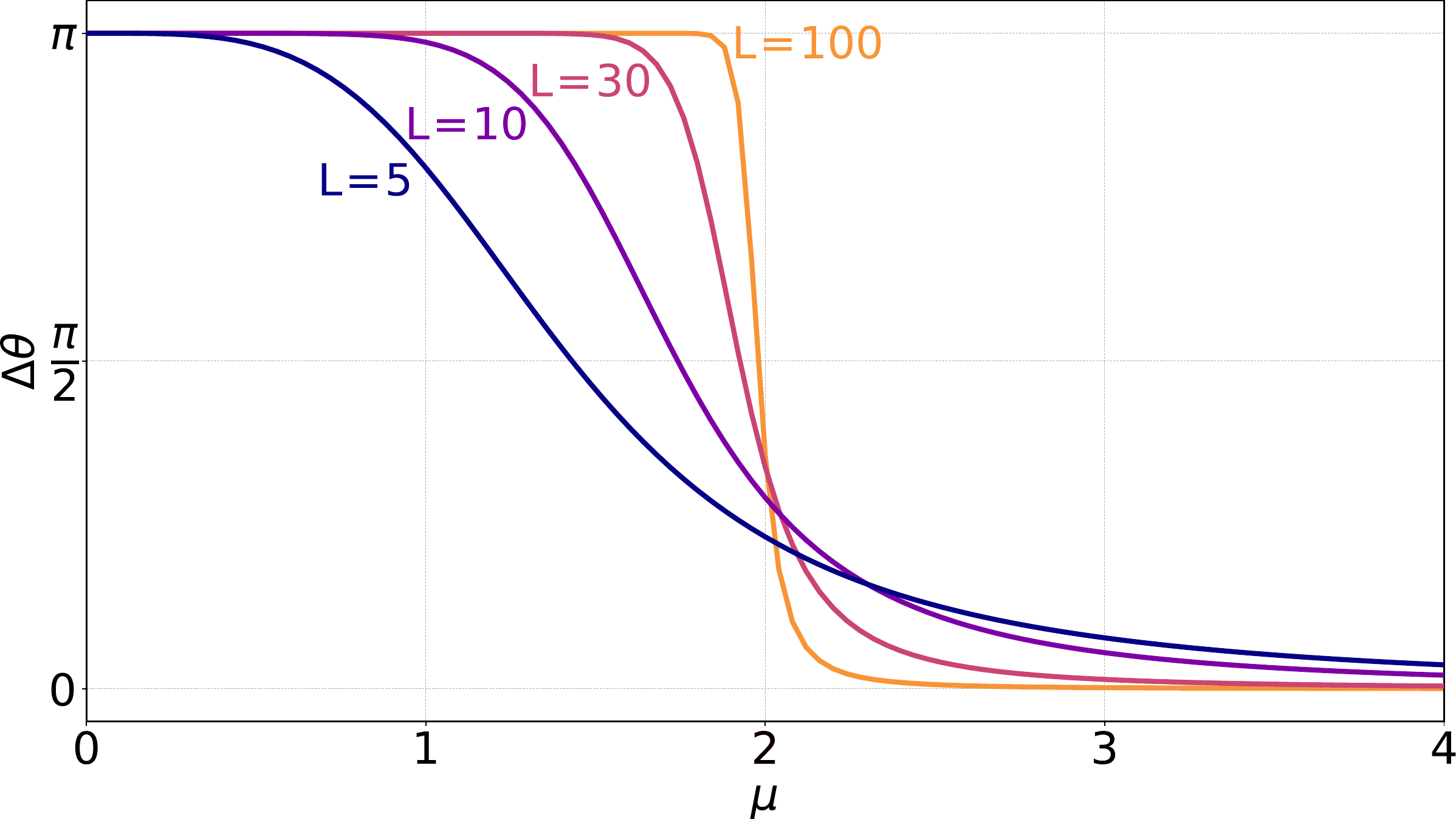}
\caption{
Difference $\Delta\theta$ between the phases acquired by $\ket{0}_{b}$ and $\ket{1}_{b}$ throughout a $2\pi$ rotation of the superconducting phase $\phi$ as a function of the chemical potential $\mu$ for different system sizes $L$. 
Topologically non-trivial phase: $\mu <2$. 
Trivial phase: $\mu> 2$.
}
\label{fig:delta_theta_over_mu}%
\end{figure}

One would expect that there is exchangeless braiding as expressed by Eq.\ (\ref{D_theta}) as long as the model supports topological Majorana boundary modes of sufficiently low energy. 
This requires that (i) the Kitaev chain remains in the non-trivial bulk topological phase and (ii) is long enough to sufficiently suppress the interaction between the MZMs.

To test this expected robustness, for parameters different from the ideal parameter set, we first consider $t = \Delta_{0}$ but $\mu =1$, such that the decay parameters $x_{\pm}$ in Eq.\ (\ref{eq:x_pm}) become non-trivial. 
Addressing point (ii) first, we have varied $L$.
As is demonstrated with Fig.\ \ref{fig:delta_Z}, the deviation of the phase difference $\Delta \theta$ from $\Delta \theta = \pi$ can be made as small as required by increasing $L$. 
Even a moderately long chain with only $L\approx 13$ sites is sufficient to get $|\Delta\theta - \pi| < 10^{-2}$.
With increasing $L$, the deviation $|\Delta\theta - \pi|$ decreases to zero exponentially.
This goes hand in hand with the exponentially decreasing energy splitting between the states $\ket{0}_{b}$ and $\ket{1}_{b}$.

To address point (i), we have fixed the chain length $L$ and varied $\mu$.
Since the nearest-neighbour hopping amplitude is set to $t=1$, the topological class of the model is completely determined by the local potential $\mu$. 
For $\mu < - 2$ and $\mu > 2$, the system (for $L\to \infty$) is topologically trivial. 
Exchangeless braiding is stable for any value $-2 < \mu < 2$, but requires ever larger system sizes when $\mu$ approaches the ($L\to \infty$) critical value $\mu_{\rm c}=2$.
This is demonstrated in Fig.\ \ref{fig:delta_theta_over_mu}, where the phase difference $\Delta \theta$ is plotted as function of $\mu$ for different fixed $L$ at $t = \Delta_{0}=1$. 
We find that the WZ phase difference gradually changes from $\Delta\theta = \pi$ to $\Delta\theta = 0$ as $\mu$ increases and passes the topological phase boundary at $\mu_{\rm c}=2$. 
This is plausible since, within the two-mode model, increasing $\mu$ eventually enhances the MZM interaction strength $h$, for any finite $L$. 
With increasing $L$, the width of the transition region decreases, and in the limit $L\to \infty$ the continuous crossover turns into a discontinuous step at $\mu_{\rm c}=2$.
Notably, Fig.\ \ref{fig:delta_theta_over_mu} also shows that even a relatively small system of $L=30$ sites supports a substantial region in which $\Delta\theta = \pi$ is stable. 

\begin{figure}[t] 
\centering
\includegraphics[width = 0.7\linewidth]{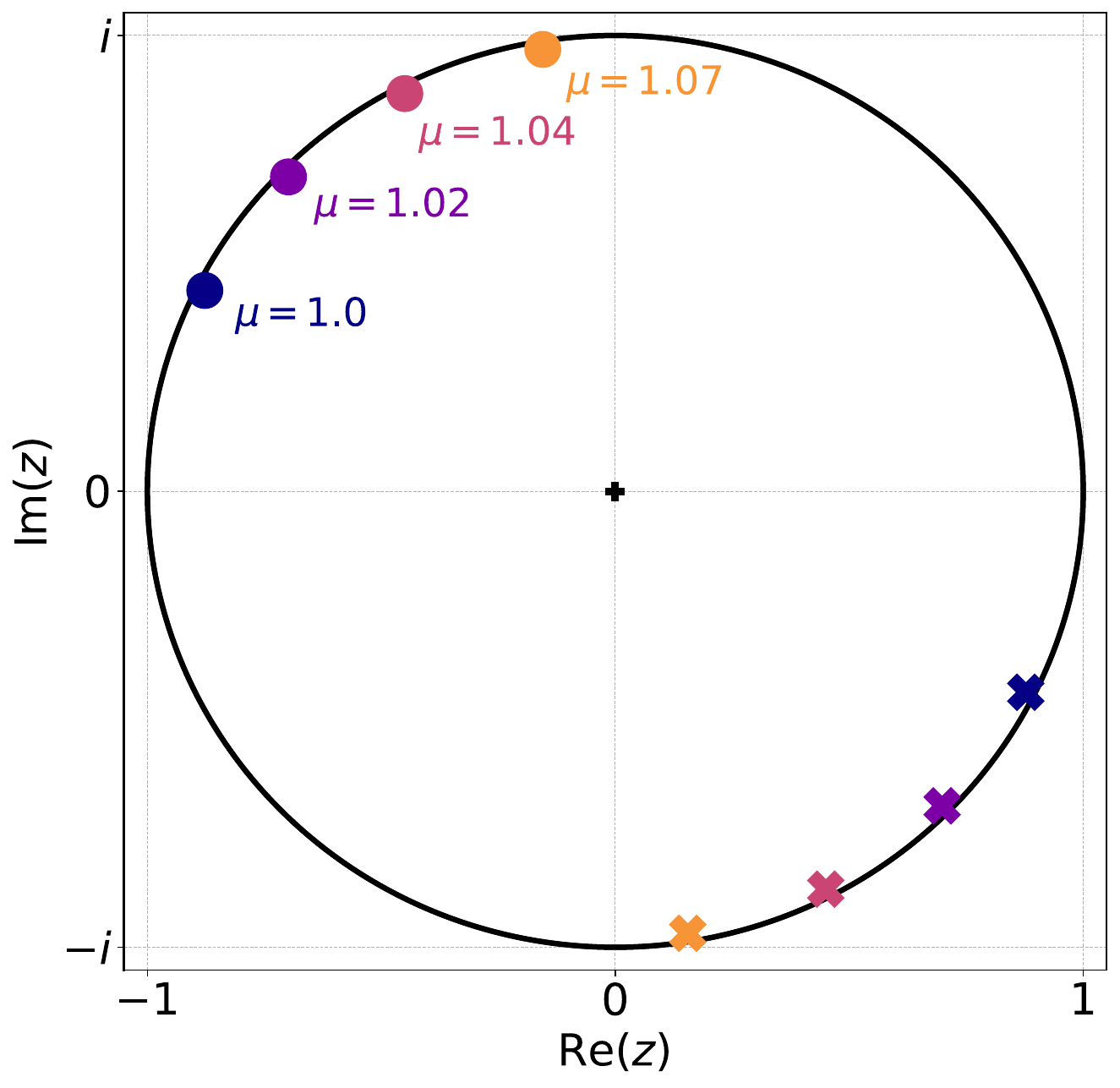}
\caption{
Phases $\theta_{0}$ (solid circles) and $\theta_{1}$ (crosses) of ground state pairs $\ket{0}_{b}$ and $\ket{1}_{b}$ for different $\mu$ as indicated in figure.
Parameters: $t = \Delta_{0} = 1.0$ and $L = 30$. 
}
\label{fig:phase_factors}
\end{figure}

Variations of $L$ and $\mu$ within that stable region only affect the global phase $\alpha$ of $U_{\text{WZ}}$ [see Eq.\ (\ref{WZ_C_1})] while preserving the anyonic phase difference of $\Delta\theta = \pi$. 
As a function of $L$, the global phase scales linearly for large $L$. 
In fact, for the Kitaev model with periodic boundaries, one can easily show analytically that $\alpha \simeq L \pi / 2$ for large $L$. 
Essentially the same behavior of $\alpha$ is also found in case of open boundaries. 
The $\mu$ dependence of the global phase close to $\mu=1$ is shown in Fig.\ \ref{fig:phase_factors} for a Kitaev chain with $\Delta_{0}=t=1$ and $L=30$.
Again, the global phase $\alpha$ is very sensitive to small changes of $\mu$ and increases linearly with increasing $\mu$.

We note that a closed path $C^{N}$ [see Eq.\ (\ref{CN})] representing $N$ full rotations of the SC phase, generates a unitary evolution on the subspace $\mathcal{H}_{0}$ that can be understood as a $2N$-fold exchangeless braiding process between $\gamma_{L}$ and $\gamma_{R}$, if the system is in the topologically non-trivial phase and sufficiently large, depending on the model parameters $t,\Delta_{0}$ and $\mu$. 

Furthermore, while a single Kitaev chain does reflect non-Abelian anyon physics, it cannot be used directly for topological quantum computing (TQC).
This is because qubits can only be realized in coherent two-dimensional Hilbert spaces and the opposite Bogoliubov quasiparticle parities of the states $\ket{0}_{b}$ and $\ket{1}_{b}$ preclude coherent superposition between them \cite{Lahtinen2017,Friis2016}. 
As a consequence, the minimal setup for TQC with Ising anyons requires four anyonic MZMs, where both the even parity sector and the odd parity sector of $\mathcal{H}_{0}$ form two-dimensional coherent subspaces capable of accommodating a topological qubit.

\section{Exchangeless Braiding in Coupled Kitaev Chains}
\label{sec:two_KC}

\begin{figure}[t] 
\centering
\includegraphics[width = 1.0\linewidth]{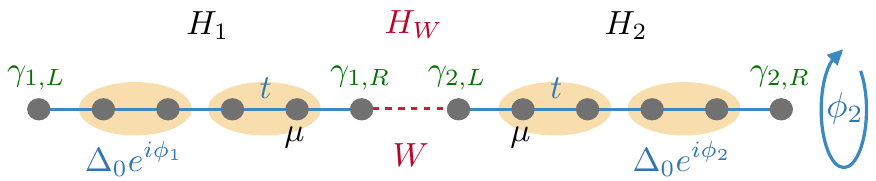}
\caption{
Two coupled Kitaev chains, see text for discussion.
}
\label{fig:chains}
\end{figure}

One way of realizing a minimal TQC system using Kitaev chains is to weakly couple two separate chains at one boundary site, see Fig.\ \ref{fig:chains}. 
We consider a system with Hamiltonian 
\be
H = H_{1} + H_{2} + H_{W}
\:, 
\label{KC_2}
\ee
where $H_{1}$ and $H_{2}$ are Kitaev-chain Hamiltonians, see Eq.\ (\ref{H_Kit}), with $L_{1}$ and $L_{2}$ sites, respectively.
We define $L = L_{1} + L_{2}$ and introduce a site index $j =1,...,L$ running over the sites of both subsystems. 
The last term in Eq.\ (\ref{KC_2}),
\be
H_{W} 
= 
- W
\left(
c_{L_{1}}^{\dagger} c_{L_{1}+1}
+
c_{L_{1}+1}^{\dagger}c_{L_{1}}
\right)
\: ,
\ee
implements a (weak) coupling between the two Kitaev subchains that links the last site $j=L_{1}$ of the first subchain to the first site $j=L_{1}+1$ of the second subchain. 
$W$ denotes the weak-link strength. 
For the rest of the paper
we use the same values for model parameters $t=1$, $\Delta_{0}$, and $\mu$ for both, $H_{1}$ and $H_{2}$ but consider independent phases $\phi_{1}$ and $\phi_{2}$ of the SC gap parameter.

In a way analogous to the single Kitaev chain, Eq.\ (\ref{KC_2}) defines a continuous family of Hamiltonians $H(\phi_{1},\phi_{2})$ but now parameterized by the two independent SC phases $\phi_{1}$ and $\phi_{2}$ of the two subchains. 
As a result, $(\phi_{1},\phi_{2})$ lives in the parameter manifold $\mathbb{S}^{1}_{\phi_{1}} \times \mathbb{S}^{1}_{\phi_{2}}\simeq \mathbb{T}^{2}_{\phi_{1},\phi_{2}}$. 
For choices of the model parameters $\Delta_{0}$ and $\mu$ that place $H_{1}$ and $H_{2}$ in the topologically non-trivial phase, the system features a total of four MZMs $\gamma_{1,L},\gamma_{1,R},\gamma_{2,L},\gamma_{2,R}$ located on the four boundary sites $j = 1, L_{1},L_{1}+1,L$ of the two subchains, respectively. 
To have well-defined and well-localized excitations, $L_{1}$ and $L_{2}$ should be large enough to sufficiently suppress the interaction of MZMs within the subchains, and $W$ should be weak enough to sufficiently suppress the inter-chain interaction of MZMs.

According to the scheme presented in Sec.\ \ref{sec:one_KC}, an exchangeless double braiding of the MZMs $\gamma_{1,L}$ and $\gamma_{1,R}$ of $H_{1}$ or $\gamma_{2,L}$ and $\gamma_{2,R}$ of $H_{2}$ can be induced by rotating the corresponding SC phase $\phi_{1}$ or $\phi_{2}$ by $2\pi$. 
In preparation for a WZ-phase-based analysis of such exchangeless braiding transformations, we extend the notion of the 
closed curves in $\mathbb{S}^{1}_{\phi}$ introduced in Eq.\ (\ref{CN}) to closed curves 
\ba
C^{N_{1},N_{2}} : [0,1] \times [0,1] 
&\to& 
\mathbb{S}^{1}_{\phi_{1}} \times \mathbb{S}^{1}_{\phi_{2}} 
\nonumber \\
(t_{1},t_{2}) 
&\mapsto& 
(2\pi N_{1} t_{1},2\pi N_{2} t_{2})
\label{C_N1_N2}
\ea
in $\mathbb{T}^{2}_{\phi_{1},\phi_{2}} = \mathbb{S}^{1}_{\phi_{1}} \times \mathbb{S}^{1}_{\phi_{2}}$. 
Here, $N_{1}$ and $N_{2}$ are the integer number of times that $C^{N_{1},N_{2}}$ winds around $\mathbb{S}^{1}_{\phi_{1}} \subset \mathbb{T}^{2}_{\phi_{1},\phi_{2}}$ and $\mathbb{S}^{1}_{\phi_{2}} \subset \mathbb{T}^{2}_{\phi_{1},\phi_{2}}$, respectively. 
In this terminology, the WZ phase along the closed curve $C^{1,0}$ measures the double exchangeless braiding of the MZMs $\gamma_{1,L}$ and $\gamma_{1,R}$ in the first subchain, while the WZ phase along the closed curve $C^{0,1}$ measures the double exchangeless braiding of the MZMs $\gamma_{2,L}$ and $\gamma_{2,R}$ in the second subchain. 
The resolution-$I$ discretization $C^{N_{1},N_{2}}_{I}$ of $C^{N_{1},N_{2}}$ is defined analogously to Eq.\ (\ref{C_I}) by restricting Eq.\ (\ref{C_N1_N2}) to the discretized domain. 
For $N_{1}=N_{2}=1$, for example, this is given by 
\be
C_{I} = \left\{\phi_{0},,...,\phi_{I-1},\phi_{I} = \phi_{0}\right\} 
\times
\left\{\phi_{0},,...,\phi_{I-1},\phi_{I} = \phi_{0}\right\} 
\: .
\label{C_II}
\ee

The presence of four MZMs makes the effect of exchangeless braiding on $\mathcal{H}_{0}$ more nuanced. 
While the single complex fermionic zero mode $b_{0} = \frac{1}{2}(\gamma_{A} + i\gamma_{B})$ in a system with two MZMs $\gamma_{A}$ and $\gamma_{B}$ is uniquely determined by them, there is an ambiguity in the definition of the two complex fermionic zero modes $b_{0,1}$ and $b_{0,2}$ supported by a system with four MZMs $\gamma_{A},\gamma_{B},\gamma_{C},\gamma_{D}$. 
For example, both
\be
b_{0,1} 
= 
\frac{1}{2} (\gamma_{A} + i \gamma_{B}) 
\;  , \quad 
b_{0,2} = \frac{1}{2} (\gamma_{C} + i \gamma_{D})
\label{b0_1_2_1}
\ee
and
\be
b_{0,1} = \frac{1}{2} (\gamma_{A} + i \gamma_{D}) 
\;  , \quad 
b_{0,2} = \frac{1}{2} (\gamma_{C} + i \gamma_{B})
\label{b0_1_2_2}
\ee
represent valid choices. This is important, because the unitary transformation $U_{AB}$ on 
\ba
\mathcal{H}_{0} 
&= &
\text{span} 
\Big\{
\ket{0} , b_{0,1}^{\dagger}\ket{0} , b_{0,2}^{\dagger}\ket{0} , b_{0,1}^{\dagger} b_{0,2}^{\dagger} \ket{0}
\Big\} 
\nonumber \\
&=& 
\text{span}
\Big\{
\ket{0,0}_{b} , \ket{1,0}_{b} , \ket{0,1}_{b} , \ket{1,1}_{b}
\Big\}
\: ,
\ea
which is induced by exchanging $\gamma_{A}$ and $\gamma_{B}$, depends strongly on the combination of the MZM operators $\gamma_{A}$, $\gamma_{B}$, $\gamma_{C}$, $\gamma_{D}$ that give the complex fermionic zero-mode operators $b_{0,1}$ and $b_{0,2}$ used in the construction of $\mathcal{H}_{0}$. 
In particular, the formal exchange of MZMs $\gamma_{A}$ and $\gamma_{B}$ results in a qualitatively different unitary transformation depending on whether $\gamma_{A}$ and $\gamma_{B}$ belong to the same complex fermion mode or not. 

This can be illustrated as follows. Consider a system with four MZMs $\gamma_{A}$, $\gamma_{B}$, $\gamma_{C}$, $\gamma_{D}$, which are paired into two complex fermionic zero-energy modes $b_{0,1}$ and $b_{0,2}$ as in Eq.\ (\ref{b0_1_2_1}). 
The unitary operator $U_{AB}$, which exchanges $\gamma_{A}$ and $\gamma_{B}$, reads
\be
U_{AB} = e^{\pi \gamma_{A}\gamma_{B}/4} = e^{i\pi/4} e^{-i\pi n_{0,1}/2}
\: ,
\label{U_AB}
\ee
where $n_{0,1} = b_{0,1}^{\dagger} b_{0,1}$ is the quasi-particle number operator associated to $b_{0,1}$. 
As is easily verified, $U_{AB}$ exchanges $\gamma_{A}$ and $\gamma_{B}$ like Ising anyons, i.e., 
\be
U_{AB} \gamma_{A} U_{AB}^{\dagger} 
= 
\gamma_{B} 
\; ,  \quad 
U_{AB} \gamma_{B} U_{AB}^{\dagger} = -\gamma_{A}
\: ,
\ee
while transforming $\gamma_{C}$ and $\gamma_{D}$ trivially. 
The matrix representation of $\mathcal{U}_{AB}$ in Eq.\ (\ref{U_AB}) with respect to basis 
$\{\ket{0,0}_{b} , \ket{1,1}_{b} , \ket{0,1}_{b} , \ket{1,0}_{b} \}$ 
constructed with the complex fermionic zero-energy mode operators $b_{0,1}$ and $b_{0,2}$ from Eq.\ (\ref{b0_1_2_1}), 
is given by 
\be
\mathcal{U}_{AB} 
= 
\begin{pmatrix}
e^{i\pi/4} & 0 & 0 & 0 \\
0 & e^{-i\pi/4} & 0 & 0 \\
0 & 0 & e^{i\pi/4} & 0 \\
0 & 0 & 0  & e^{-i\pi/4}
\end{pmatrix}
\: .
\label{U_AB_mat}
\ee
A double exchange of $\gamma_{A}$ and $\gamma_{B}$ is therefore described by the diagonal U(4) transformation
\be
\mathcal{U}_{AB}^{2} 
= e^{i\pi/2} 
\begin{pmatrix}
1 & 0 & 0 & 0 \\
0 & -1 & 0 & 0 \\
0 & 0 & 1 & 0 \\
0 & 0 & 0  & -1
\end{pmatrix}
\: .
\label{U_AB_mat_sq}
\ee

If instead the four MZMs $\gamma_{A}$, $\gamma_{B}$, $\gamma_{C}$, $\gamma_{D}$ are paired into two complex fermionic zero-energy modes $b_{0,1}$ and $b_{0,2}$ as in Eq.\ (\ref{b0_1_2_2}), the unitary operator $U_{AB}$ exchanging $\gamma_{A}$ and $\gamma_{B}$ becomes
\ba
U_{AB} 
&=& 
e^{\pi \gamma_{A}\gamma_{B}/4}  
\nonumber \\
&=& 
\frac{1}{\sqrt{2}} \left(1+ i(b_{0,1}^{\dagger}+b_{0,1}) (b_{0,2}^{\dagger}-b_{0,2})\right)
\: .
\label{U_AB_2}
\ea
The matrix representation of $\mathcal{U}_{AB}$ in Eq.\ (\ref{U_AB_2}) 
with respect to the same basis $\{\ket{0,0}_{b},\ket{1,1}_{b},\ket{0,1}_{b},\ket{1,0}_{b}\}$ reads
\be
\mathcal{U}_{AB} 
= 
\frac{1}{\sqrt{2}}\begin{pmatrix}
1  & \makeupminus{i}  & 0  & 0  \\
\makeupminus{i}  & 1  & 0  & 0  \\
0  & 0 & 1  & -i  \\
0 & 0  & -i  & 1
\end{pmatrix}
\: ,
\ee
and it differs from Eq.\ (\ref{U_AB_mat}), because this time it is based on the definition for $b_{0,1}$ and $b_{0,2}$ in Eq.\ (\ref{b0_1_2_2}) and thus on alternative complex fermionic zero-energy modes.
As a consequence, a double exchange of $\gamma_{A}$ and $\gamma_{B}$ in the new basis of $\mathcal{H}_{0}$ corresponds to the non-diagonal U(4) transformation
\be
\mathcal{U}_{AB}^{2} 
= 
\begin{pmatrix}
0 & \makeupminus{i} & 0 & 0 \\
\makeupminus{i} & 0 & 0 & 0 \\
0 & 0 & 0 & -i \\
0 & 0 & -i  & 0
\end{pmatrix}
\: .
\label{U_AB_mat_sq_alt}
\ee
Thus, the way in which the four MZMs combine into two complex fermionic zero modes determines the effect of their braiding on the complex subspace $\mathcal{H}_{0}$ of degenerate many-body ground states. 

For a fixed chain length $L$ and for fixed $\Delta_{0} = t = 1$, the ratio between the weak-link strength $W$ and the local potential $\mu$ controls the relative strength of inter- and intra-subchain interactions between MZMs. 
In the following, we show that this relative strength in turn determines how the MZMs combine into the complex fermionic low-energy modes. 
Specifically, dominant intra-chain interaction leads to pairings
\be
b_{0,1} 
= 
\frac{1}{2} (\gamma_{1,L} + i \gamma_{1,R}) 
\, , \quad
b_{0,2} = \frac{1}{2} (\gamma_{2,L} + i \gamma_{2,R})
\: ,
\label{b0_1_2_teaser1}
\ee
while dominant inter-chain interaction instead produces
\be
b_{0,1} 
= 
\frac{1}{2} (\gamma_{1,L} + i \gamma_{2,R}) 
\: ,  \quad 
b_{0,2} = \frac{1}{2} (\gamma_{1,R} + i \gamma_{2,L})
\: .
\label{b0_1_2_teaser2}
\ee
Varying the ratio between $W$ and $\mu$ therefore allows us to tune between Eq.\ (\ref{b0_1_2_teaser1}) and Eq.\ (\ref{b0_1_2_teaser2}), so that we can choose whether the exchangeless braiding between $\gamma_{s,L}$ and $\gamma_{s,R}$ induced by a $2\pi$ rotation of the SC phase in the $s$-th subchain ($s=1,2$) results in a braiding transformation of the form Eq.\ (\ref{U_AB_mat_sq}) or Eq.\ (\ref{U_AB_mat_sq_alt}).

We first discuss the trivial case of two disconnected chains, i.e., $W=0$. 
Here, the MZMs of either subchain must combine into a complex fermion mode locally, and hence the complex fermionic zero modes are of the form Eq.\ (\ref{b0_1_2_teaser1}). 
A $2\pi$ phase rotation in either subchain induces an exchangeless braiding between MZMs from the same complex fermion, giving a U(4) WZ phase matrix of the form given by Eq.\ (\ref{U_AB_mat_sq}) resembling a $Z$-gate.

For a finite but sufficiently  weak link $W>0$ between the two subchains, we expect essentially the same result.
In a generic system with subchains of different lengths, $L_{1} \neq L_{2}$, there are two interaction strengths $h_{1}$ and $h_{2}$ associated with the intra-subchain interactions between the MZMs of either subchain, see Eqs.\ (\ref{eq:Maj_int}) -- (\ref{eq:x_pm}). 
It appears plausible that the $Z$-gate-type braiding must then persist with increasing $W$ until 
\be
W \simeq h_{\text{max}} \equiv \max\{h_{1},h_{2}\}
\: ,
\label{4mode}
\ee
where $h_{1}$ and $h_{2}$ are each obtained from Eqs.\ (\ref{eq:Maj_int_h}), (\ref{eq:int_length}), and (\ref{eq:x_pm}).
This gives us a four-mode theory for two coupled Kitaev chains as a straightforward generalization of the two-mode theory discussed in Sec.\ \ref{sec:one_KC}. 
Compared to the full theory of Sec.\ \ref{sec:meth} and as we will see below, the simplified four-mode model turns out to be quite accurate.

\begin{figure}[t] 
\centering
\includegraphics[width = 1.0\linewidth]{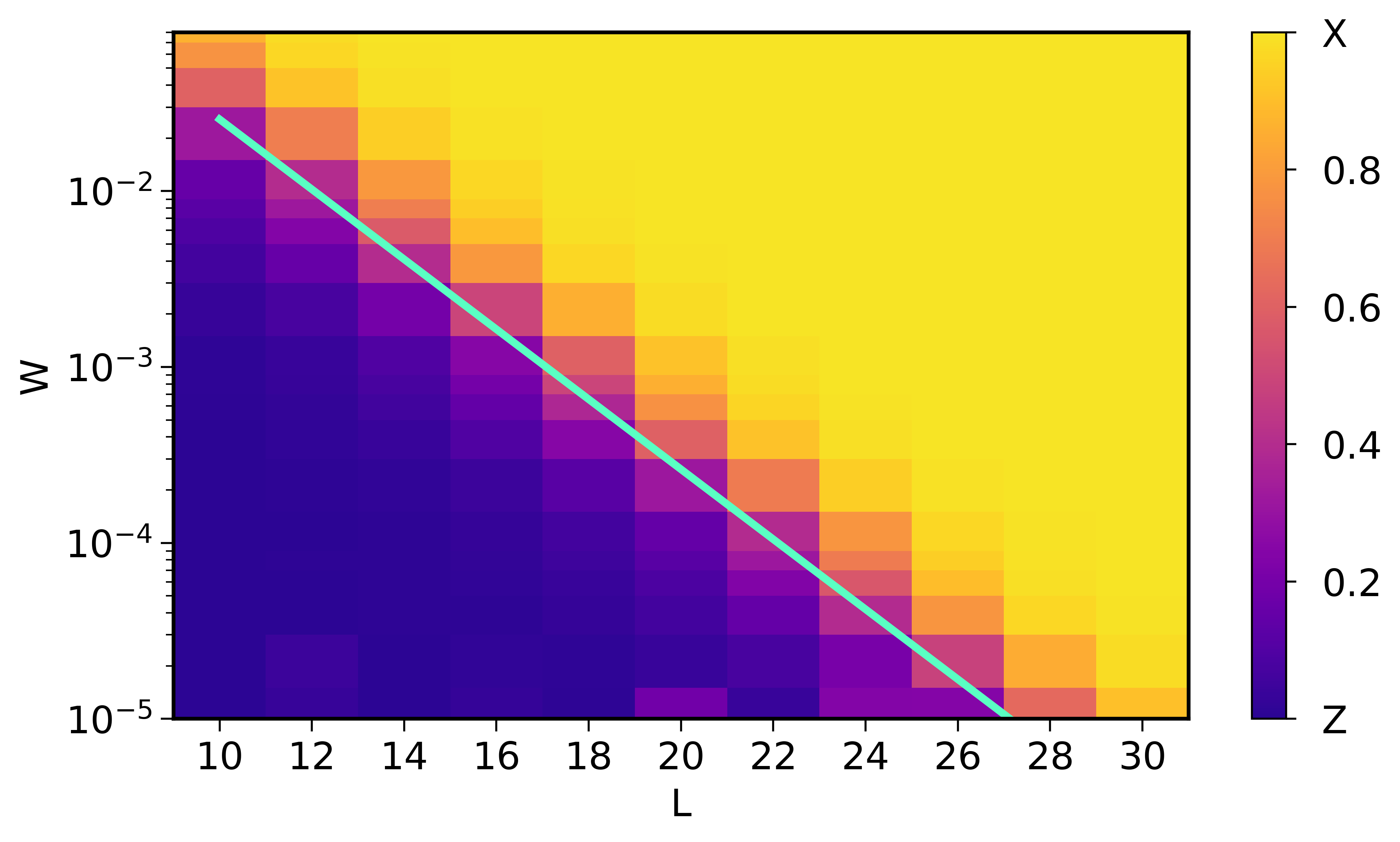}
\caption{
Phase diagram showing the braiding type of the WZ phase matrix as a function of the weak-link strength $W$ and the total system size $L$ for $\Delta_{0}=t=1$ and $\mu=0.8$. 
The color code quantifies $d=d(U_{\text{WZ}}[C^{0,1}])$, see Eqs.\ (\ref{distz}), (\ref{distx}), (\ref{distd}) and the related discussion. 
$d=0$ for a Z-type WZ phase matrix, Eq.\ (\ref{WZ_num_2}), and $d=1$ for an X-type matrix, Eq.\ (\ref{WZ_num_3_Xgate}).
The cyan line $W=h_{\max}$, Eq.\ (\ref{4mode}), is the result of the simple four-mode model (see text).
}
\label{fig:PD_LW}
\end{figure}

Numerical calculations have been carried out for systems of different sizes $L$. 
We have taken $L$ to be even and assumed $L_{1} = L/2 - 1$ and $L_{2} = L/2 + 1$ for convenience. 
This slight spatial asymmetry helps to avoid unwanted degeneracies. 
The dependence of $W = h_{\text{max}}$ on the system size $L$, as obtained from Eqs.\ (\ref{4mode}), (\ref{eq:Maj_int_h}) -- (\ref{eq:x_pm}), is shown as the cyan line in Fig.\ \ref{fig:PD_LW} for $\Delta_{0}=t=1$ and $\mu=0.8$. 
Already for very small systems with $L=14$ (i.e., $L_{1} =6$, $L_{2}=8$), for example, we have $h_{\text{max}} \lesssim 10^{-2}$. As expected,
$h_{\rm max}$ decreases exponentially with increasing $L$. 

Opposed to the four-mode model, the full theory does not predict a sharp transition between the Z-gate phase at weak $W$ and small $L$, and the X-gate phase at strong $W$ and large $L$, but rather a smooth crossover. 
The results for $\Delta_{0}=t=1$ and $\mu=0.8$, shown in Fig.\ \ref{fig:PD_LW} (see color code and discussion below), are obtained by approximating the WZ phase $U_{\text{WZ}}[C^{0,1}]$ along the closed curve $C^{0,1}$ from Eq.\ (\ref{C_N1_N2}) by its discretized version with a resolution of $I = 1000$. 
This has proven fully sufficient to obtain converged results. 

We first discuss the results for the extreme limits: 
Deep in the Z-gate phase and for small $L$ and weak $W$, see Fig.\ \ref{fig:PD_LW}, we find a WZ phase matrix of the form
\be
U_{\text{WZ}}[C^{0,1}] 
\approx e^{i\theta} \begin{pmatrix}
1 & 0 & 0 & 0\\
0 & -1 & 0 & 0 \\
0 & 0 & 1 & 0\\
0 & 0 & 0 & -1 \\
\end{pmatrix}
\equiv Z
\; ,
\label{WZ_num_2}
\ee
with respect to the basis $\{\ket{0,0}_{b},\ket{1,1}_{b},\ket{0,1}_{b},\ket{1,0}_{b}\}$. 
Up to an irrelevant global parameter-dependent phase factor $e^{i\theta}$, this matches the expected result Eq.\ (\ref{U_AB_mat_sq}) for a system of disconnected subchains. 
Restricting the WZ phase matrix Eq.\ (\ref{WZ_num_2}) to the even and odd parity sectors $\mathcal{H}_{0}\vert_{P_{0}}$ and $\mathcal{H}_{0}\vert_{P_{1}}$ of the degenerate subspace $\mathcal{H}_{0}$ yields
\be
U_{\text{WZ}}[C^{0,1}] \vert_{P_{0}} = U_{\text{WZ}}[C^{0,1}] \vert_{P_{1}} \approx e^{i\theta} \begin{pmatrix}
1 & 0 \\
0 & -1
\end{pmatrix}
\: ,
\label{U2_PZ}
\ee
where $U_{\text{WZ}}[C^{0,1}] \vert_{P_{0}}$ and $U_{\text{WZ}}[C^{0,1}] \vert_{P_{1}}$ are given with respect to the bases $\{\ket{0,0}_{b},\ket{1,1}_{b}\}$ and $\{\ket{0,1}_{b},\ket{1,0}_{b}\}$, respectively. 
As for $W=0$, see Eq.\ (\ref{D_theta}), we find for the phase difference
\be
\Delta\theta\big(U_{\text{WZ}}[C^{0,1}] \vert_{P_{0}}\big)
=
\Delta\theta\big(U_{\text{WZ}}[C^{0,1}] \vert_{P_{0}}\big)
\approx 
\pi
\: .
\ee
By definition, the parity sectors of $\mathcal{H}_{0}$ describe two-level systems of equal-parity states. As a consequence, they can be used to encode a qubit and the projective $\sigma_{z}$ transformations in Eq.\ (\ref{U2_PZ}) can be understood as $Z$-quantum gates on that qubit.

For strong $W$ and large $L$, i.e., deep in the $X$-gate phase, our numerical calculations yield
\be
U_{\text{WZ}}[C^{0,1}] 
\approx 
e^{i\theta} 
\begin{pmatrix}
0 & e^{-i\alpha} & 0 & 0\\
e^{i\alpha} & 0 & 0 & 0 \\
0 & 0 & 0 & e^{-i\beta}\\
0 & 0 & e^{i\beta} & 0 \\
\end{pmatrix}
\: ,
\label{WZ_num_3}
\ee
with respect to the basis $\left\{\ket{0,0}_{b},\ket{1,1}_{b},\ket{0,1}_{b},\ket{1,0}_{b}\right\}$. 
Up to a global parameter-dependent phase factor $e^{i\theta}$ and after a suitable gauge-transformation
\be
U_{\text{WZ}}[C^{0,1}] 
\mapsto 
G \,  U_{\text{WZ}}[C^{0,1}] \, G^{\dagger}
\ee
with a unitary matrix $G \in T < \text{U}(4)$ of the form 
\be
G 
= 
\text{diag}(e^{i\gamma_{1}},e^{i\gamma_{2}},e^{i\gamma_{3}},e^{i\gamma_{4}})
\: ,
\label{G_gauge}
\ee
$U_{\text{WZ}}[C^{0,1}]$ in Eq.\ (\ref{WZ_num_3}) is equivalent to the expected result Eq.\ (\ref{U_AB_mat_sq_alt}). 
This can, for example, be achieved by the projective global phase $\theta=\pi/2$ in combination with the gauge transformation characterized by $\gamma_{1} = - \gamma_{2} = \alpha / 2$ and $\gamma_{3} = - \gamma_{4} = (\beta-  \pi) / 2$. 
A similar calculation shows that Eq.\ (\ref{WZ_num_3}) is also equivalent to the braiding transformation
\be
U_{\text{WZ}}[C^{0,1}] 
\approx
e^{i\theta}  
\begin{pmatrix}
0 & 1 & 0 & 0\\
1 & 0 & 0 & 0 \\
0 & 0 & 0 & 1 \\
0 & 0 & 1 & 0 \\
\end{pmatrix}
\equiv 
X
\: .
\label{WZ_num_3_Xgate}
\ee
Thus, the restrictions of the WZ phase matrix in Eq.\ (\ref{WZ_num_3}) to the even and odd parity sectors $\mathcal{H}_{0}\vert_{P_{0}}$ and $\mathcal{H}_{0}\vert_{P_{1}}$ of $\mathcal{H}_{0}$ are both equivalent to
\be
U_{\text{WZ}}[C^{0,1}] \vert_{P_{0}} 
= 
U_{\text{WZ}}[C^{0,1}] \vert_{P_{1}} 
\simeq 
e^{i\theta} 
\begin{pmatrix}
0 & 1\\
1 & 0
\end{pmatrix}
\: ,
\label{U2_PX}
\ee
where $U_{\text{WZ}}[C^{0,1}] \vert_{P_{0}}$ and $U_{\text{WZ}}[C^{0,1}] \vert_{P_{1}}$ are given in the bases $\{\ket{0,0}_{b},\ket{1,1}_{b}\}$ and $\{\ket{0,1}_{b},\ket{1,0}_{b}\}$, respectively. 
For strong $W$ and large $L$, the unitary transformation induced by exchangeless double braiding between the MZMs of one subchain can therefore be understood as an $X$-gate on the equal-parity sectors of $\mathcal{H}_{0}$.

\begin{figure}[t] 
\centering
\includegraphics[width = 1.0\linewidth]{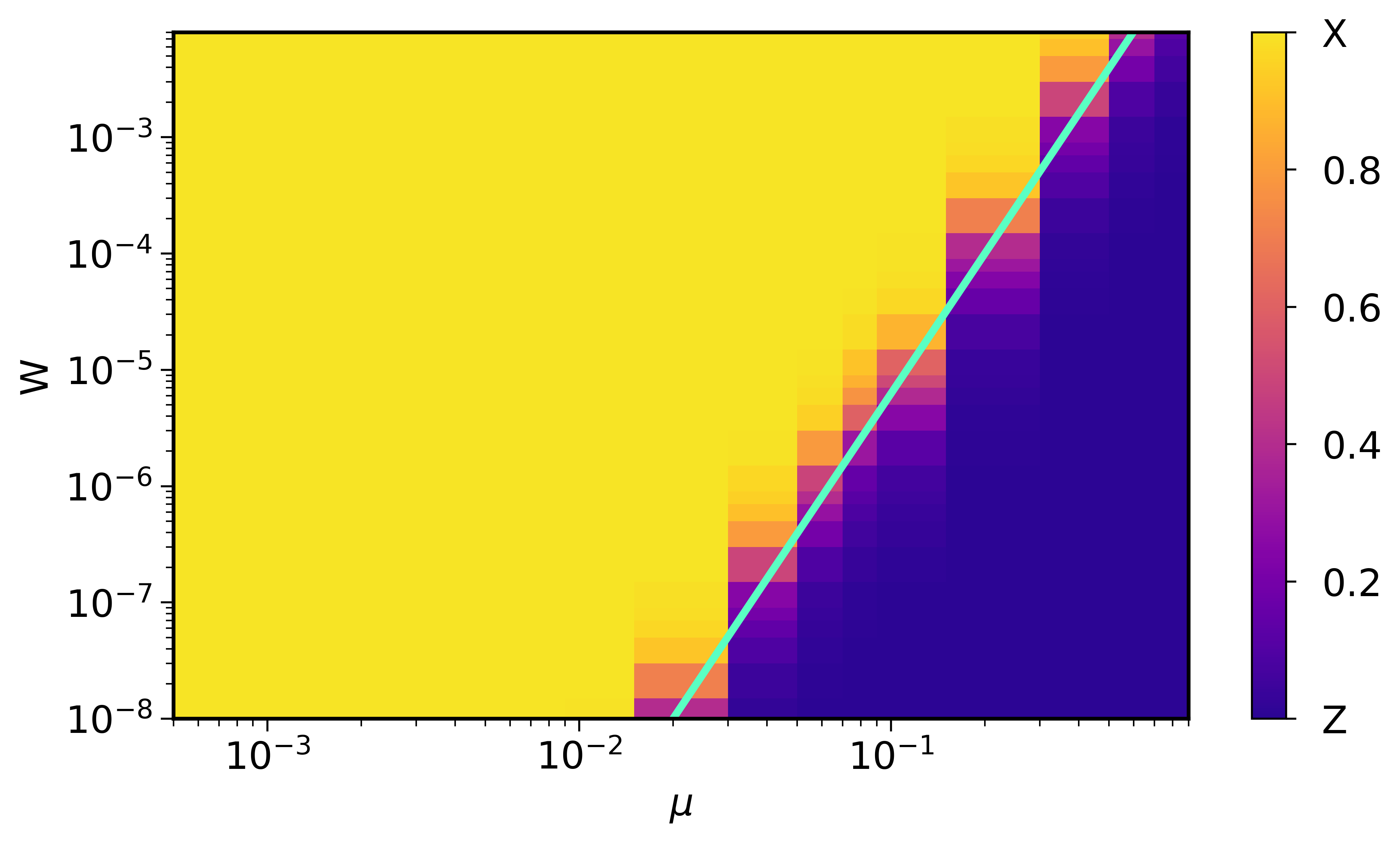}
\caption{
The same as in Fig.\ \ref{fig:PD_LW} but showing the braiding type of the WZ phase matrix as a function of $W$ and $\mu$, at fixed $\Delta_{0}=t=1$ and $L=10$.
}
\label{fig:PD_VW}
\end{figure}

Numerical results for the intermediate parameter regimes are shown as an $L$-$W$ phase diagram in Fig.\ \ref{fig:PD_LW} and as a $\mu$-$W$ phase diagram in Fig.\ \ref{fig:PD_VW}. 
For the construction of the phase diagrams, we use the Frobenius norm $\| A \| = \sqrt{\mbox{tr}(A^{\dagger} A)}$ of a matrix $A$ to define the distance of a WZ phase matrix $U = U_{\text{WZ}}[C^{0,1}]$ from $Z$ [see Eq.\ (\ref{WZ_num_2})], modulo phase factors, as 
\be
d_{Z}(U) = \big\| |U| - |Z| \big\|
\label{distz}
\ee
and the distance from $X$ [see Eq.\ (\ref{WZ_num_3_Xgate})] as 
\be
d_{X}(U) = \big\|  |U| - |X|  \big\|
\label{distx}
\: .
\ee
Here, $|A|$ is the matrix with elements $|A_{ij}|$.
With this, we can quantify the crossover from the $Z$-gate to the $X$-gate phase via
\be
d(U) = \frac{d_{Z}(U)} { \sqrt{d_{X}^{2}(U) + d_{Z}^{2}(U)} }
\: .
\label{distd}
\ee
We have $d(U_{\text{WZ}}[C^{0,1}]) = 0$ deep in the Z-gate phase, and
$d(U_{\text{WZ}}[C^{0,1}]) = 1$ deep in the X-gate phase. 
It has been verified that in both, the deep Z- and X-gate phases, the WZ phase matrix, {\em including the phase factors of the matrix elements}, approaches the form given by Eqs.\ (\ref{WZ_num_2}) and (\ref{WZ_num_3_Xgate}), respectively. 

As can be seen in Figs.\ \ref{fig:PD_LW} and \ref{fig:PD_VW}, the transition between the $Z$- and $X$-type braiding WZ phase matrices involves a crossover region, in which the braiding type of the WZ phase matrix is not defined. 
In both figures, the region is quite accurately centered around the transition line $W=h_{\text{max}}$ of the four-mode model (cyan lines).
Moving away from the $h_{\rm max}$ line, either to small $L$, weak $W$, large $\mu$ or to large $L$, strong $W$, small $\mu$, 
i.e., leaving the finite crossover regime, the WZ phase matrix quickly approaches the $Z$- or $X$-gate form, respectively. 

To realize $Z$ or $X$ quantum gates, small systems with $L=\ca O (10^{1})$ sites are sufficient. 
At $L=12$ and $\mu = 0.8$, see Fig.\ \ref{fig:PD_LW} for example, one can switch between Z and X by switching the weak-link strength between $W<10^{-3}$ and $W>10^{-1}$. 
At $L=10$ and $\mu=0.2$, see Fig.\ \ref{fig:PD_VW} for example, a tiny change from $W \sim 10^{-6}$ to $W \sim 10^{-2}$ induces a transition from Z to X gates.
Note that, while our discussion refers to numerical results obtained for the choice $\Delta_{0} = t$, this choice is just convenient [see Eq.\ (\ref{eq:x_pm})], and our conclusions do not at all rely on this. 
The location of the crossover parameter regime itself is reliably described by the four-mode model.
The width of the crossover, however, is essential as well and can be captured with the full many-body theory only.

\section{Conclusions}
\label{sec:conclusion}

Majorana zero modes (MZMs) of topological Kitaev chain networks were shown to largely retain their anyonic properties despite weak hybridizations caused by finite chain lengths or coupling between chains. 
This makes it possible to realize exchangeless braiding protocols of MZMs even in networks of short and weakly coupled Kitaev chains and opens up access to a parameter regime with tunable braiding results. 
In particular, we have studied two simple Kitaev networks and used $2\pi N$ rotations of the superconducting phase $\phi$ to induce exchangeless braiding protocols. 
An analysis of these protocols was based on the non-Abelian Wilczek-Zee (WZ) phase of the low-energy many-body subspaces $\mathcal{H}_{0}(\phi)$. 
The main technical component for the numerical evaluation of the WZ phase is the Bertsch-Robledo formula Eq.\ (\ref{overlap}), which allows an efficient computation of many-body overlaps and only requires numerically stable Bogolyubov diagonalization and Bloch-Messiah decomposition algorithms. 

A single Kitaev chain of finite length was considered first.
In the topological phase, the model supports two boundary MZMs, which combine into a single complex Bogoliubov quasi-particle mode of (nearly) zero energy, giving rise to a two-dimensional subspace $\mathcal{H}_{0}$ of (nearly) degenerate many-body ground states. 
The MZMs of the topological Kitaev chain are known to behave as Ising anyons and are predicted to undergo an effective double exchange when the superconducting phase $\phi$ is advanced by $2 \pi$ \cite{Kitaev2001,Chiu2016}. 
Using the full many-body framework, we could confirm this prediction numerically by computing the $\text{U}(2)$ WZ phase of $\mathcal{H}_{0}$ along $2 \pi$ rotations of $\phi$ and by showing that it is projectively equivalent to the square of the Ising anyon exchange matrix. 
In particular, we quantify the resilience of the anyon properties of the MZMs in the presence of finite-size induced hybridization. 
It is found that even short chains of, e.g., $L \approx 30$ sites support anyon physics over most of the topological phase given by 
$|\mu| < 2|t|$.

We furthermore investigated a minimal network of two weakly-coupled Kitaev chains of finite lengths. 
The introduction of a second chain brings three independent interaction strengths into the system: 
the strength $W$ of the weak link connecting the two Kitaev chains, and the two emergent hybridization strengths $h_{1}$ and $h_{2}$ that result from the finite lengths of the individual chains. 
This added complexity enables a transition between two regions in parameter space, each characterized by the outcome of the exchangeless braiding transformation induced by rotating the superconducting phase $\phi_{i}$ of either subchain. 
Given that all interaction strengths are weak compared to the bare energy scales of the system, we found that dominant weak-link interactions produce exchangeless braiding transformations resembling a projective $\sigma_{x}$ gate. 
Contrary, a projective $\sigma_{z}$ gate is obtained for dominant intra-subchain interactions.
The transition between the two regions is a continuous crossover around a line predicted by a simple four-mode model and given by $W=h_{\text{max}}(L,\mu)$, where $h_{\text{max}}$ denotes the stronger of the two intra-subchain interactions $h_{1}$ and $h_{2}$.

The crossover regime deserves special attention, since here the Wilczek-Zee phase matrix deviates from an ideal X- or Z-gate phase matrix, i.e., in this parameter regime exchangeless braiding lacks strict topological protection.
In fact, this is a rather universal phenomenon resulting from a finite overlap of edge modes and affecting all topological systems of finite size. 
For small systems, where this overlap is not exponentially small in the system size, it is very important to quantify the extent of this crossover parameter regime. 

Our study is based on the analysis of the non-Abelian WZ geometrical phase associated with the continuous family of Kitaev-chain Hamiltonians parameterized by the phase of the superconducting gap. 
Besides this important geometrical property, the dynamical contributions to the non-Abelian phase in an adiabatic process can be taken into account additionally in a subsequent step to improve the predictive power for physical realizations of robust quantum gates.
In addition, the formal framework used here can also be applied to study time-dependent observables in a real-time simulation, as has been demonstrated recently \cite{Mascot2023}. 
This allows studying effects of real-time dynamics beyond the adiabatic limit, e.g., non-adiabatic MZM braiding dynamics \cite{Sanno2021}.

Addressing the full real-time quantum-classical dynamics \cite{EP21,LLP22,QP22} of more realistic systems represents another important future step. 
Here, an interesting example is given by a system where local magnetic moments, modelled as classical spins, are locally exchange-coupled to a conventional BCS $s$-wave superconductor. 
The low-energy electronic structure emerging from the interaction of in-gap Yu-Shiba-Rusinov states is predicted to resemble Kitaev-chain physics, and one expects that global rotations of the classical-spin configuration are equivalent to global rotations of the phase of the $p$-wave superconducting gap that is proximity induced in the Shiba bands \cite{Pientka2013}.
In such quantum-classical models, the classical degrees of freedom used to control the exchangeless braiding process are themselves dynamic variables.

\acknowledgments

This work was supported by the Deutsche Forschungsgemeinschaft (DFG, German Research Foundation) through the research unit QUAST, FOR 5249 (project P8), Project ID No.\ 449872909, and through the Cluster of Excellence “Advanced Imaging of Matter” - EXC 2056, Project ID No.\ 390715994. 

\section*{Data availability}

The data that support the findings of this article are openly available \cite{data}.

\end{document}